\documentclass[a4paper]{article}
\usepackage{epsfig}
\usepackage{graphicx}  % standard LaTeX graphics tool
\usepackage{color}     % for color figures

\usepackage{color}
\usepackage{lscape}
\date{\today}
 \normalsize

\newcommand{\bmat}{\left(\begin{array}}
\newcommand{\emat}{\end{array}\right)}
\newcommand{\be}{\begin{equation}}
\newcommand{\ee}{\end{equation}}
\newcommand{\bea}{\begin{eqnarray}}
\newcommand{\eea}{\end{eqnarray}}

\def\gtwid{\mathrel{\raise.3ex\hbox{$>$\kern-.75em\lower1ex\hbox{$\sim$}}}}
\def\ltwid{\mathrel{\raise.3ex\hbox{$<$\kern-.75em\lower1ex\hbox{$\sim$}}}}
\def\gev{{\rm \, Ge\kern-0.125em V}}
\def\tev{{\rm \, Te\kern-0.125em V}}

\def\B{{\bf B}}

\def    \be            {\begin{equation}}
\def    \ee            {\end{equation}}
\def    \bea           {\begin{eqnarray}}
\def    \eea           {\end{eqnarray}}

\def\d{\delta}
\def\n{\nu}

\def\th{\theta}

\def\d{\delta}
\def\D{\Delta}
\def\s{\sigma}
\def\r{\rho}
\def\t{\theta}
\def\e{\epsilon}

\def\nn{\nonumber}
\begin{document}
\renewcommand{\thefootnote}{\fnsymbol{footnote}}
%\rightline{IPPP/03/52} \rightline{DCPT/03/104}
\vspace{.3cm}

\title{\Large\bf Zero minors of the neutrino mass matrix  }

\author
{ \it \bf  E. I. Lashin$^{1,2,3}$\thanks{elashin@ictp.it} and N.
Chamoun$^{1,4}$\thanks{nchamoun@hiast.edu.sy} ,
\\ \small$^1$ The Abdus Salam ICTP, P.O. Box 586, 34100 Trieste, Italy. \\
\small$^2$ Ain Shams University, Faculty of Science, Cairo 11566,
Egypt.
\\ \small$^3$ Department of physics and Astronomy, College
of Science, King Saud University, Riyadh, Saudi Arabia,\\
\small$^4$  Physics Department, HIAST, P.O.Box 31983, Damascus,
Syria.  }

\maketitle

\begin{center}
\small{\bf Abstract}\\[3mm]
\end{center}
We examine the possibility that a certain class of neutrino mass
matrices, namely those with two independent
vanishing minors in the flavor basis, regardless of being invertible or not,
is sufficient to describe current data.
We compute generic formulae for the ratios of the neutrino masses and
for the Majorana phases. We find that seven textures with two
vanishing minors  can accommodate the experimental data. We
present an estimate of the mass matrix for these patterns. All
the possible textures can be dynamically generated through the seesaw
mechanism augmented with a discrete Abelian symmetry.

\vspace{1.1cm}{\bf PACS numbers}: 14.60.Pq; 11.30.Hv; 14.60.St
\begin{minipage}[h]{14.0cm}
\end{minipage}
\vskip 0.3cm \hrule \vskip 0.5cm
%%%%%%%%%%%%%%%%%%%%%%%%%%%%%%%%%%
\section{Introduction}
The results of the Super-Kamiokande \cite{SK} on the solar
neutrino ($\nu_e$) deficit and the atmospheric neutrino
($\nu_\mu$) anomaly can be interpreted if attributed to neutrino
oscillations, which in turn can naturally occur if neutrinos are
massive and lepton flavors are mixed \cite{Bilenky}. Massive
neutrinos are commonly believed to be Majorana particles, and at
low energy scales, the phenomenology of lepton masses and flavor
mixing can be formulated in terms of the charged lepton mass
matrix $M_l$ and the (effective) neutrino mass matrix $M_\nu$.
They totally involve twelve physical parameters: three masses of
charged leptons ($m_e$, $m_\mu$ and $m_\tau$), which have
precisely been measured \cite{PDG02}; three masses of neutrinos
($m_1$, $m_2$ and $m_3$), whose relative sizes have been estimated
from solar and atmospheric neutrino oscillations
\cite{SNO,KM,K2K}; three angles of flavor mixing ($\theta_x$,
$\theta_y$ and $\theta_z$), whose values have been constrained to
an acceptable accuracy also from solar, atmospheric and reactor
neutrino oscillations \cite{SNO,KM,K2K,CHOOZ}; and three phases of
CP violation (two Majorana-type $\rho$, $\sigma$ and one
Dirac-type $\delta$), which are completely unrestricted by current
neutrino data.

One can work in the flavor basis which identifies the flavor
eigenstates of the charged leptons with their mass eigenstates, so
that $M_\nu$ will contain nine free parameters. However, in order
to account for the experimental constraints, we need extra
assumptions for $M_\nu$. The general idea is to assume that some
independent matrix elements of $M_\nu$ are actually dependent upon
one another, caused by an underlying (broken) flavor symmetry. In
particular, this dependence becomes very simple and transparent,
if the relevant matrix elements are exactly equal to zero.

In fact, general categories of zero-textures were studied: Out of
the twenty possible patterns of $M_\n$ with three independent
vanishing entries, none is allowed by current neutrino oscillation
data, while of the fifteen possible patterns of $M_\nu$ with two
independent vanishing entries, there are nine patterns which are
found to be compatible with current experimental data (albeit two
of them are only marginally allowed \cite{FGM,Xing}). A detailed phenomenological study for the
two-textures zero was given in \cite{kumar}. As to the
six possible one-zero textures of $M_\nu$, a phenomenological analysis in the general case was given in \cite{merle}, whereas
a specific model having two heavy right handed neutrinos in
was suggested in \cite{Xing03b} led  to a vanishing mass
eigenvalue.

% and there are three textures of $M_\nu$ with $m_1 =0$
%and four textures of $M_\n$ with $m_3 =0$ which are compatible
%with the current neutrino data .

In this paper, we adopt a new texture of  two independent
vanishing minors in the mass matrix $M_\nu$. Since a zero-element
can be viewed as a zero-determinant of a $1 \times 1$ sub-matrix,
then the pattern with vanishing minors (which are determinants of
$2 \times 2$ sub-matrices) can be viewed as a generalization of
the zero-textures belonging to the same category of matrices with
vanishing sub-determinants. In fact, a zero minor in $M_\n$ can be
related to a zero entry in the Majorana mass matrix of the right
handed singlet neutrinos $M_R$ in the canonical see-saw mechanism:
\bea
M_\n &=& M_D M_R^{-1} M_D^T, \label{see-saw}
\eea
where $M_D$
is the Dirac neutrino mass matrix. It was argued \cite{masub} that
the zeros of $M_R$ have a deeper theoretical meaning than the
texture zero of $M_\n$ and that if $M_D$ is diagonal then texture
zeros of $M_R$ are reflected in $M_\n$ as zero minors. To keep
$M_D$ diagonal and to maintain the form of the studied pattern of
$M_\n$, a suitable family symmetry ($A_4$) was introduced in
\cite{maa4}.

The work of \cite{Lavoura} investigated viable textures with two
zeros in the inverted neutronino mass matrix $M_\nu ^{-1}$ and,
using abelian symmetries with one or two heavy scalar singlets,
realizations of these textures were constructed. Our work assuming
two vanishing minors is somehow different from the work of
\cite{Lavoura} in the following three points:
\begin{itemize}
\item It is true that a vanishing cofactor of the neutrino mass matrix
implies a zero in the inverse neutrino mass matrix,
but the equivalence comes when the inverse exists. Differently
from \cite{Lavoura}, we did not assume that the neutrino mass
matrix is invertible, and in some cases in our work we could make
one of the masses $m_1$ or $m_3$ exactly equal to zero.
%%%%%%%%%%%%%%%%%%%%%%%%%%%%%%%%%%%%%%%%%%%%%%%%%%%%%%%
\item The
motivation for studying the texture of vanishing minors is somehow
different form that for studying the zero texture in the inverse
mass matrix, in the sense that we are generalizing the zero
texture in a non trivial way, regardless of its relation to zero
textures in other matrices. In \cite{SayedTrace}, textures
containing two independent traceless $2\times 2$ sub-matrices were
studied, while here we study textures of two independent
`determinant-less' $2\times 2$ sub-matrices.
%%%%%%%%%%%%%%%%%%%%%%%%%%%%%%%%%%%%%%%%%%%%%%%%%%%%%%
\item The phenomenological analysis in this paper contains different
details from that of \cite{Lavoura}. The strategy we followed
consisted of the fact that putting two minors equal to zero gives
us four real conditions, so with plausible values for five given
input data parameters (the three mixing angles $\theta_x$,
$\theta_y$ and $\theta_z$, and the Dirac CP-violating phase
$\delta$, and one input taken to be the solar neutrino
mass-squared difference $\Delta m^2_{\mbox{sol}}$), one should be able to
test the validity of the model to fit the other data.
We have varied the known neutrino parameters in their allowed ranges and checked whether vanishing minors
are allowed.
%We have
%spanned almost the whole parameter space in order to find
%phenomenologically acceptable regions.
Whereas in
\cite{SayedTrace}, eight textures out of the  possible fifteen
textures of vanishing two-subtraces were shown to be allowed by
experimental data, here we find that seven zero-minor textures, divided into three categories, are
able to accommodate the current data, with three patterns showing only normal type hierarchy. One, and only one, acceptable
two-vanishing minor texture of singular neutrino mass matrix can be obtained, when $m_3=0$ and $\theta_z =0$, in three patterns
which, otherwise, are failing.
%    onlyand five others can
%marginally do so, while three cases fail completely.
\end{itemize}

In accordance with \cite{Lavoura}, we give an order of
magnitude to the neutrino mass matrix, for acceptable values of the input parameters, and find that four acceptable patterns of the
two-zeroes textures can be reproduced, with analysis conforming to \cite{Xing}.

%Of these latter six cases,
%four patterns, already studied in , are allowed
%phenomenologically. This classification of patterns agrees well
%with the work of .

The plan of the paper is as follows: in section $2$, we review
the standard notation for the three-flavor neutrino oscillations
and its relation to the experimental constraints. In section $3$,
we present the texture of $M_\nu$ with two independent vanishing
minors and compute the expressions of the two neutrino mass ratios
and the Majaorana phases. We classify the patterns and present the results and the phenomenological
analysis of each case in
sections $4$--$8$. The symmetry realization of all models is presented
in section $9$. We end up by conclusions in section $10$.

\section{Standard notation}

In the flavor basis where the charged lepton mass matrix
is diagonal, the symmetric neutrino mass matrix
$M_\nu$ can be diagonalized by a unitary transformation,
\begin{equation}
V^{\dagger} M_\nu\; V^{*} \; = \; \left (\matrix{ m_1 & 0 & 0 \cr 0 & m_2 & 0
\cr 0 & 0 & m_3 \cr} \right ), \;
\end{equation}
with $m_i$ (for $i=1,2,3$) real and positive, while the lepton
flavor mixing matrix $V$ can be written as a product of a
Dirac-type flavor mixing matrix $U$ (consisting of three mixing
angles and one CP-violating phase) and a diagonal phase matrix $P$
(consisting of two nontrivial Majorana phases): $V = UP$ where $
P=\mbox{diag}(e^{i\rho},e^{i\sigma},1)$. With
\begin{equation}
\lambda_1 \; =\; m_1 e^{2i\rho} \; , ~~~ \lambda_2 \; =\; m_2
e^{2i\sigma} \; , ~~~ \lambda_3 = m_3, \; \end{equation}  we may
rewrite $M_\nu$ as
\begin{equation}
M_\nu \; =\; U \left ( \matrix{ \lambda_1 & 0 & 0 \cr 0 &
\lambda_2 & 0 \cr 0 & 0 & \lambda_3 \cr} \right ) U^T. \;
\label{mnu}
\end{equation}

As to the matrix $U$, and taking the indices ($1,2,3$) to refer to
the flavors ($e,\mu,\tau$) respectively, it can be parameterized
as \cite{Xing}:

\begin{equation}
U \; = \; \left ( \matrix{ c_x c_z & s_x c_z & s_z \cr - c_x s_y
s_z - s_x c_y e^{-i\delta} & - s_x s_y s_z + c_x c_y e^{-i\delta}
& s_y c_z \cr - c_x c_y s_z + s_x s_y e^{-i\delta} & - s_x c_y s_z
- c_x s_y e^{-i\delta} & c_y c_z \cr } \right ) \; ,
%   (11)
\end{equation}
where $s_x \equiv \sin\theta_x$, $c_x \equiv \cos\theta_x$ (we
will use later also the notation $t_x$ for $\tan x$), and so on.
From eq.(\ref{mnu}), the mass matrix elements
take the forms:
\bea
M_{\n\,11}&=& m_1 c_x^2 c_z^2 e^{2\,i\,\r} + m_2 s_x^2 c_z^2 e^{2\,i\,\s}
+ m_3\,s_z^2,\nn\\
%%%%%%%%%%%%%%%%%%%%%%%%%%%%%%%%%%%%%
M_{\n\,12}&=& m_1\left( - c_z s_z c_x^2 s_y e^{2\,i\,\r}
- c_z c_x s_x c_y e^{i\,(2\,\r-\d)}\right)
+ m_2\left( - c_z s_z s_x^2 s_y e^{2\,i\,\s}
+ c_z c_x s_x c_y e^{i\,(2\,\s-\d)}\right) + m_3 c_z s_z s_y,\nn\\
%%%%%%%%%%%%%%%%%%%%%%%%%%%%%%%%%%
M_{\n\,13}&=& m_1\left( - c_z s_z c_x^2 c_y e^{2\,i\,\r}
+ c_z c_x s_x s_y e^{i\,(2\,\r-\d)}\right)
+ m_2\left( - c_z s_z s_x^2 c_y e^{2\,i\,\s}
- c_z c_x s_x s_y e^{i\,(2\,\s-\d)}\right) + m_3 c_z s_z c_y,\nn\\
%%%%%%%%%%%%%%%%%%%%%%%%%%%%%%%%%
M_{\n\,22}&=& m_1 \left( c_x s_z s_y  e^{i\,\r}
+ c_y s_x e^{i\,(\r-\d)}\right)^2 + m_2 \left( s_x s_z s_y  e^{i\,\s}
- c_y c_x e^{i\,(\s-\d)}\right)^2 + m_3 c_z^2 s_y^2, \nn\\
%%%%%%%%%%%%%%%%%%%%%%%%%%
M_{\n\, 23} &=& m_1\left( c_x^2 c_y s_y s_z^2  e^{2\,i\,\r}
 + s_z c_x s_x (c_y^2-s_y^2) e^{i\,(2\,\r-\d)} - c_y s_y s_x^2 e^{2\,i\,(\r-\d)}\right)
\nn\\
&& +  m_2\left( s_x^2 c_y s_y s_z^2  e^{2\,i\,\s}
 + s_z c_x s_x (s_y^2-c_y^2) e^{i\,(2\,\s-\d)} - c_y s_y c_x^2 e^{2\,i\,(\s-\d)}\right)
+ m_3 s_y c_y c_z^2, \nn\\
%%%%%%%%%%%%%%%%%%%%%%%%%%%%%%%%%%%%%%%%%%%%
M_{\n\,33}&=& m_1 \left( c_x s_z c_y  e^{i\,\r}
- s_y s_x e^{i\,(\r-\d)}\right)^2 + m_2 \left( s_x s_z c_y  e^{i\,\s}
+ s_y c_x e^{i\,(\s-\d)}\right)^2 + m_3 c_z^2 c_y^2.
\label{melements}
\eea
If we denote by $\e$ to the transposition on the set ${1,2,3}$ which swaps the two indices $2$ and $3$, then
under the symmetry $T$ on the parameters defined by : \begin{equation} \label{T}
T\hspace{12pt}:\hspace{12pt}\t_y\rightarrow
\frac{\pi}{2}-\t_y, \delta\rightarrow\delta\pm \pi
\end{equation} we have \bea \label{invariance} \left( M_{\n}\right)_{i,j} (T(\t_y),T(\d))&=& \left( M_{\n
}\right)_{\e(i),\e(j)} (\th_y,\d
).\eea
This fact can be, gainfully, used for distinguishing
the subcategories of the neutrino mass matrices within the same class.

A
remarkable merit of this parametrization is that its three mixing
angles $(\theta_x, \theta_y, \theta_z)$ are directly related to
the mixing angles of solar, atmospheric and CHOOZ reactor neutrino
oscillations:
\begin{equation}
\theta_x \; \approx \; \theta_{\rm sun} \; , ~~~~~ \theta_y \;
\approx \; \theta_{\rm atm} \; , ~~~~~ \theta_z \; \approx \;
\theta_{\rm chz}. \;
\end{equation}
Also we have,
\begin{equation}
\Delta m^2_{\mbox{sol}} \; = \;
\Delta m^2_{12}=m_2^2-m_1^2 \; , \;\Delta m^
2_{\mbox{atm}}  \;
%\approx |\Delta m^2_{31}|
= \; |\Delta m^2_{23}|= \left|m_3^2-m_2^2\right|\;\; ,
\end{equation}
and the hierarchy of solar and
atmospheric neutrino mass-squared differences is characterized by
the parameter:
\begin{equation}
R_\nu \; \equiv \; \left | \frac{m^2_2 - m^2_1} {m^2_3 - m^2_2}
\right | \; \approx \; \frac{\Delta m^2_{\rm sun}} {\Delta
m^2_{\rm atm}} \; \ll \; 1 \; .
%   (9)
\end{equation}
Reactor nuclear experiments on beta-decay kinematics and
neutrinoless double-beta decay put constraints on the neutrino
mass scales characterized by the following parameters: the
effective electron-neutrino mass
\begin{equation}
\langle
m\rangle_e \; = \; \sqrt{\sum_{i=1}^{3} \displaystyle \left (
|V_{e i}|^2 m^2_i \right )} \;\; ,
%       (2.23)
\end{equation} and the effective Majorana mass term
$\langle m \rangle_{ee} $ given by
\begin{equation} \langle m \rangle_{ee} \; = \; \left | m_1
V^2_{e1} + m_2 V^2_{e2} + m_3 V^2_{e3} \right | \; .
%       (2.30)
\end{equation}
Also, cosmological observations put an upper bound on the `sum'
parameter $\Sigma$ which is:
\be \Sigma = \sum_{i=1}^{3} m_i.\ee

A recent global analysis of neutrino experimental data (\cite{fog}
and references therein), gives the following  estimates, at the
confidence level of $95\%$, for the above parameters:
%%%%%%%%%%%%%%%%%%%%%%%%%%%%%%%%%%%%%%%%%%%%%%%%%%%%%
\bea
\Delta m^2_{\mbox{atm}} &=& \left( 2.4^{+0.5}_{-0.6}\right) \times
10^{-3}\;\mbox{eV}^2,
\nonumber \\
\Delta m^2_{\mbox{sol}} &=& \left( 7.92\pm 0.7 \right)
\times 10^{-5}\;\mbox{eV}^2, \nonumber\\
\sin^2\theta_{\mbox{atm}} = \left(
0.44^{+0.18}_{-0.01}\right) &\longleftrightarrow & \th_y = \left(
41.55^{+10.4}_{-5.6}\right)\mbox{degree}, \nonumber \\
\sin^2\theta_{\mbox{sol}} = \left( 0.314^{+0.057}_{-0.047}\right)
&\longleftrightarrow & \th_x = \left(
34.08^{+3.4}_{-3}\right)\mbox{degree}, \nonumber\\
\sin^2\theta_{\mbox{chz}} = \left( 0.9^{+2.3}_{-0.9}\right)\times 10^{-2}
&\longleftrightarrow & \th_z = \left(
5.44^{+5}_{-5}\right)\mbox{degree},\nonumber\\
\langle m\rangle_e &<& 1.8\; \mbox{eV}, \nonumber \\
\Sigma &<& 1.4 \;\mbox{eV}, \nonumber\\
\langle m \rangle_{ee} &=& \left(
0.58^{+0.22}_{-0.16}\right) \mbox{eV}.
\label{expdata}
\eea
In particular, the bounds on
$R_\n$ put very stringent conditions on any model required to fit
the data:
\bea \label{Rconstraint} R_\nu &=& \left(
0.033^{+0.016}_{-0.008}\right).
\eea
Note that the lower bound on
$\langle m \rangle_{ee}$ disappears if the neutrinoless
double-beta decay does not exist. Moreover, this value is obtained based on the `claimed' observation
of \cite{Klapdor}, which is not unanimously agreed upon.

\section{Neutrino mass matrices with two vanishing minors}
As $M_\nu$ is $3 \times 3$ symmetric matrix, it totally has $6$
independent complex entries, and thus it has $6$ independent
minors. We will denote by $C_{ij}$ the minor corresponding to the
$ij^{th}$ element (i.e. the determinant of the sub-matrix obtained
by deleting the $i^{th}$ row and the $j^{th}$ column of $M_\nu$).
Hence, we have $15$ possibilities of having vanishing two minors.

Following the classification of \cite{FGM,Lavoura}, we list in table \ref{modelnames} the fifteen possible two-
vanishing minor texture with its defining minors.
\begin{table}[tb]
\begin{center}
\begin{tabular}{|c|c|}
\hline
 Pattern  &        Vanishing minors         \\
 \hline
 $A_1$ &     $C_{33}$,\, $C_{32}$     \\
 $A_2$ &    $C_{22}$,\, $C_{32}$      \\
 $B_3$ &    $C_{33}$,\, $C_{31}$  \\
 $B_4$ &   $C_{22}$,\, $C_{21}$ \\
 $B_5$ &    $C_{33}$,\, $C_{12}$   \\
 $B_6$ &  $C_{22}$,\, $C_{13}$  \\
 $D$   &  $C_{33}$,\, $C_{22}$  \\
 $S_1$ &   $C_{31}$,\, $C_{11}$  \\
 $S_2$ & $C_{21}$,\, $C_{11}$ \\
 $S_3$ & $C_{13}$,\, $C_{12}$ \\
 $F_1$ & $C_{33}$,\, $C_{11}$ \\
 $F_2$ & $C_{22}$,\, $C_{11}$ \\
 $F_3$ & $C_{32}$,\, $C_{11}$ \\
 $F_4$ & $C_{31}$,\, $C_{32}$ \\
 $F_5$ & $C_{21}$,\, $C_{32}$ \\
 \hline
\end{tabular}
\end{center}
\caption{ Vanishing two-minors texture
 mass matrices.}
\label{modelnames}
\end{table}
The patterns $A_1$, $A_2$, $B_3$, $B_4$, $S_1$ and $S_2$ are two-texture zeros \cite{Lavoura}, and the first four
of them can accommodate the data \cite{FGM} (In \cite{FGM}, there were other three acceptable two-texture zero denoted by
$B_1$, $B_2$ and $C$ which are not reproduced as vanishing two-minor texture). The symmetry $T$ in eqs. (\ref{T}, \ref{invariance})
 transforms the neutrino mass matrices of patterns ($A_1$, $B_3$, $B_5$, $S_1$, $F_1$ and $F_4$) to ($A_2$, $B_4$, $B_6$, $S_2$, $F_2$ and $F_5$)
 respectively, whereas the patterns ($D$, $S_3$ and $F_3$) are singlets under $T$.
Therefore, the predictions,  for parameters ($\t_y, \,\d$), of one pattern in the first set are the same as its corresponding pattern
in the second set, for parameters ($\frac{\pi}{2}-\t_y, \, \d \pm \pi$).

If two minors vanish, we have \bea
\label{det1} M_{\nu\;ab}\;M_{\nu\;cd} - M_{\nu\;ij}\; M_{\nu\;mn} &=& 0,  \\ \label{det2}
M_{\nu\;a'b'}\;M_{\nu\;c'd'} - M_{\nu\;i'j'}\; M_{\nu\; m'n'} &=& 0,
\eea
then we have
\bea
\label{U's} \sum_{l,k=1}^{3}\left(
U_{al}U_{bl}U_{ck}U_{dk}-U_{il}U_{jl}U_{mk}U_{nk} \right)
\lambda_l \lambda_k &=& 0,
\eea
and a similar equation with
$(abcdijmn)$ replaced by their `primes'. We get
\bea
\frac{\lambda_1}{\lambda_3} &=& \frac{K_2 L_1 - K_1 L_2}{K_2 L_3 -
K_3 L_2},  \\
\frac{\lambda_2}{\lambda_3} &=& \frac{K_2 L_1 - K_1
L_2}{K_1 L_3 - K_3 L_1},
\eea
where
\bea
\label{Ah}
K_h&=& \left(
U_{al}U_{bl}U_{ck}U_{dk}-U_{il}U_{jl}U_{mk}U_{nk} \right) + \left(
l\leftrightarrow k \right), \\
L_h&=&\left(
U_{a'l}U_{b'l}U_{c'k}U_{d'k}-U_{i'l}U_{j'l}U_{m'k}U_{n'k} \right)
+ \left( l \leftrightarrow k \right),
\eea
with ($h,l,k$) are a
cyclic permutation of ($1,2,3$).

In this way, with the input of four parameters determining the
matrix $U$ (the three mixing angles $\theta_x$, $\theta_y$,
$\theta_z$ and the Dirac phase $\delta$), we are able to predict
the relative magnitude of the three neutrino masses and the values
of the two Majorana phases from the relations:
\bea
\frac{m_i}{m_3}&=&\left| \frac{\lambda_i}{\lambda_3}\right|
\,\,\mbox{for}\,\,i=1,2,
\eea
and
\bea
\rho &=& \frac{1}{2}\; \mbox{arg} \left( \frac{\lambda_1}{\lambda_3} \right),\\
\sigma &=& \frac{1}{2}\; \mbox{arg} \left( \frac{\lambda_2}{\lambda_3}
\right).
\eea
We can examine now whether or not the chosen texture
of $M_\nu$ is empirically acceptable by computing the magnitude of
the parameter $R_\nu$ which should be in the order of $10^{-2}$
(equation \ref{Rconstraint}). With some plausible values of the
input parameters, and taking $\Delta m^2_{\mbox{sol}}$ to be its
`central' allowable experimental value, one can reconstruct the
mass matrix and test whether or not the other experimental
constraints are respected.

We found that the resulting mass patterns could be classified into
two categories:
\begin{itemize}
\item Normal hierarchy: characterized by $m_1\sim m_2 < m_3$ and
is denoted by ${\bf N}$. \item Inverted hierarchy: characterized
by $m_1\sim m_2 > m_3$ and is denoted by ${\bf I}$.
\end{itemize}
%In addition, there are three patterns which are not suitable to
%accommodate the data.

The possibility of having non-invertible mass matrix is examined
for each pattern. The viable non-invertible mass matrices are
characterized by vanishing one of the masses ($m_1, \;
\mbox{and} \; m_3$), as compatibility with the data prevents $m_2$ to vanish. The conditions and relations
satisfied for each possibility are as follows:
\begin{itemize}
\item The vanishing of $m_1$ implies that $K_1=L_1=0$ for the same set of parameters, and the mass
spectrum of $m_2$ and $m_3$ takes the values $\sqrt{\Delta
m^2_{\mbox{sol}}}$ and $\sqrt{\Delta m^2_{\mbox{sol}}+\Delta
m^2_{\mbox{atm}}}$   respectively.

 \item The vanishing
of $m_3$ implies that $K_3=L_3=0$ and the mass spectrum of $m_2$
and $m_1$  takes the values $\sqrt{\Delta m^2_{\mbox{atm}}}$ and
$\sqrt{\Delta m^2_{\mbox{atm}}- \Delta m^2_{\mbox{sol}}}$
respectively.
\end{itemize}

%\section{Results of textures with two vanishing minors}
We present in the following sections the analysis of all the patterns corresponding to two vanishing
minors, referred to by their corresponding elements, in the
neutrino mass matrix $M_\nu$. When the expressions are complicated
we only state the analytical leading terms of the expansions in
powers of $s_z$. The numerical estimates, for quick reference, are shown in table
(\ref{tab1}), where we fixed, when possible, the input
parameters ($\th_x,\th_y,\th_z$) to their `experimental' centered
values ($34^\circ, 42^\circ, 5^\circ$).

\section{Class A}

%%%%%%%%%%%%%%%%%%%%%%%%%%%%%%%%%%%%%%%%%% A1  %%%%%%%%%%%%%%%%%%%%%%%%%%%%%%%%%%%%%
%%%%%%%%%%%%%%%%%%%%%%%%%%%%%%%%%%%%%%%%%%%%%%%%%%%%%%%%%%%%%%%%%%%%%%%%%%%%%%%%%%%%%%%%

 {\it Pattern} {\bf A1}: vanishing minors ($C_{33}$,\, $C_{32}$): We get

\bea
\label{xingA1-1}\frac{\lambda_1}{\lambda_3} &=&
\frac{s_z}{c_z^2}\left(\frac{s_xs_y}{c_xc_y}e^{i\d}-s_z\right),
 \nonumber \\
 \frac{\lambda_2}{\lambda_3} &=&
 -\frac{s_z}{c_z^2}\left(\frac{c_xs_y}{s_xc_y}e^{i\d}+s_z\right).
 \eea

This texture is the two-texture zero neutrino mass matrix having zeros at the ($1,1$) and ($1,2$) entries \cite{Lavoura}. Consequently,
 it implies the absence of neutrinoless double-beta
decay:
\bea
\langle m \rangle_{ee} &=& 0
\eea
and thus can have only normal type hierarchy \cite{jenkins}.
We have the following analytical approximations:
\bea
\label{xingA1-2}
\frac{m_1}{m_3} &=& t_x t_y s_z - c_\d s_z^2+ O \left(
{s_z}^{3} \right),\nonumber \\
\frac{m_2}{m_3} &\approx& \frac{t_y}{t_x} s_z + c_\d s_z^2 +O \left(s_z^3\right), \nonumber \\
\frac{m_2}{m_3} - \frac{m_1}{m_3} &=& t_y (\frac{1}{t_x} - t_x) s_z + 2 c_\d s_z^2 +O \left(s_z^3\right), \nonumber \\
R_\nu &=& \left | \frac {4t_y^2}{s_{2x}t_{2x}} {s_z}^{2} \right| + O \left(
{s_z}^{3} \right),\nonumber\\
\rho &=& \frac{\delta}{2} + \frac{s_\d s_z}{2 t_x t_y}   +O \left( s_z^2\right) \left( \mbox{mod}\, \frac{\pi}{2} \right), \nonumber \\
 \sigma &=& \frac{\delta}{2}- \frac{s_\d t_x s_z}{2 t_y} +O \left( s_z^2\right) \left( \mbox{mod}\, \frac{\pi}{2} \right),\nonumber \\
 \r - \s &=& \frac{s_\d s_z}{2 t_y} (t_x + \frac{1}{t_x})+O \left( s_z^2\right) \left( \mbox{mod}\, \frac{\pi}{2} \right),\nn \\
 \frac{\langle
m \rangle_{e}}{m_3} &\approx& \frac{1}{c_y} s_z+O \left( {s_z}^{2}
 \right).
 \eea

Taking the constraints eq. (\ref{expdata}) into consideration, we see that no quasi-degenerate spectrum can be obtained,
 since we have $m_2/m_3 < 0.4$. We also note that there is a lower bound $\t_z > 4.52^\circ$, from the expression of $R_\n$ and
 the maximal and minimal values the
 parameters can take, and that maximal mixing ($\t_y = \frac{\pi}{4}$) is allowed.
The parameter $\d$ is not experimentally constrained and satisfying the $R_\n$-constraint (eq. \ref{Rconstraint}) does not single out specific
values for it.
We matched the the data with  $(\t_x = 34^0, \t_y = 42^0, \d = 0^0, \t_z = 7^0)$.
For these inputs we obtain $m_1/m_3 = 0.060$, $m_2/m_3 = 0.180$,
$\r  = 0^0$, $\s = 90^0$ and $R_\nu = 0.0298$. The mass $m_3$
fitted from the observed $\D m^2_{\mbox{sol}}$ is $m_3 = 0.052\;
\mbox{eV}$, and the other values for the remaining parameters are
$\D m^2_{\mbox{atm}}=2.7\times 10^{-3}\;\mbox{eV}^2$, $\langle m \rangle_e
=0.009\;\mbox{eV}$, $\langle m \rangle_{ee} =0 \;\mbox{eV}$ and $\Sigma =0.065
\;\mbox{eV} $. In this pattern the numerically estimated mass
matrix $M$ is \be M_\nu= m_3\,\left(
\begin {array}{ccc}
0.363\times 10^{-8}-0.228\times 10^{-10}\,i&-0.90\times 10^{-9}-0.234\times 10^{-10}\,i
& 0.165+0.248\times 10^{-10}\,i\\
0.90\times 10^{-9}-0.234\times 10^{-10}\,i& 0.396-0.241\times 10^{-10}\,i
&0.543+0.255\times 10^{-10}\,i\\
 0.165+0.248\times 10^{-10}\,i&0.543+0.255\times 10^{-10}\,i& 0.483-0.271\times 10^{-10}\,i
\end {array}
\right).
\label{massA1}
\ee

\begin{figure}[hbtp]
\centering
\begin{minipage}[c]{0.5\textwidth}
\epsfxsize=5.5cm
\centerline{\epsfbox{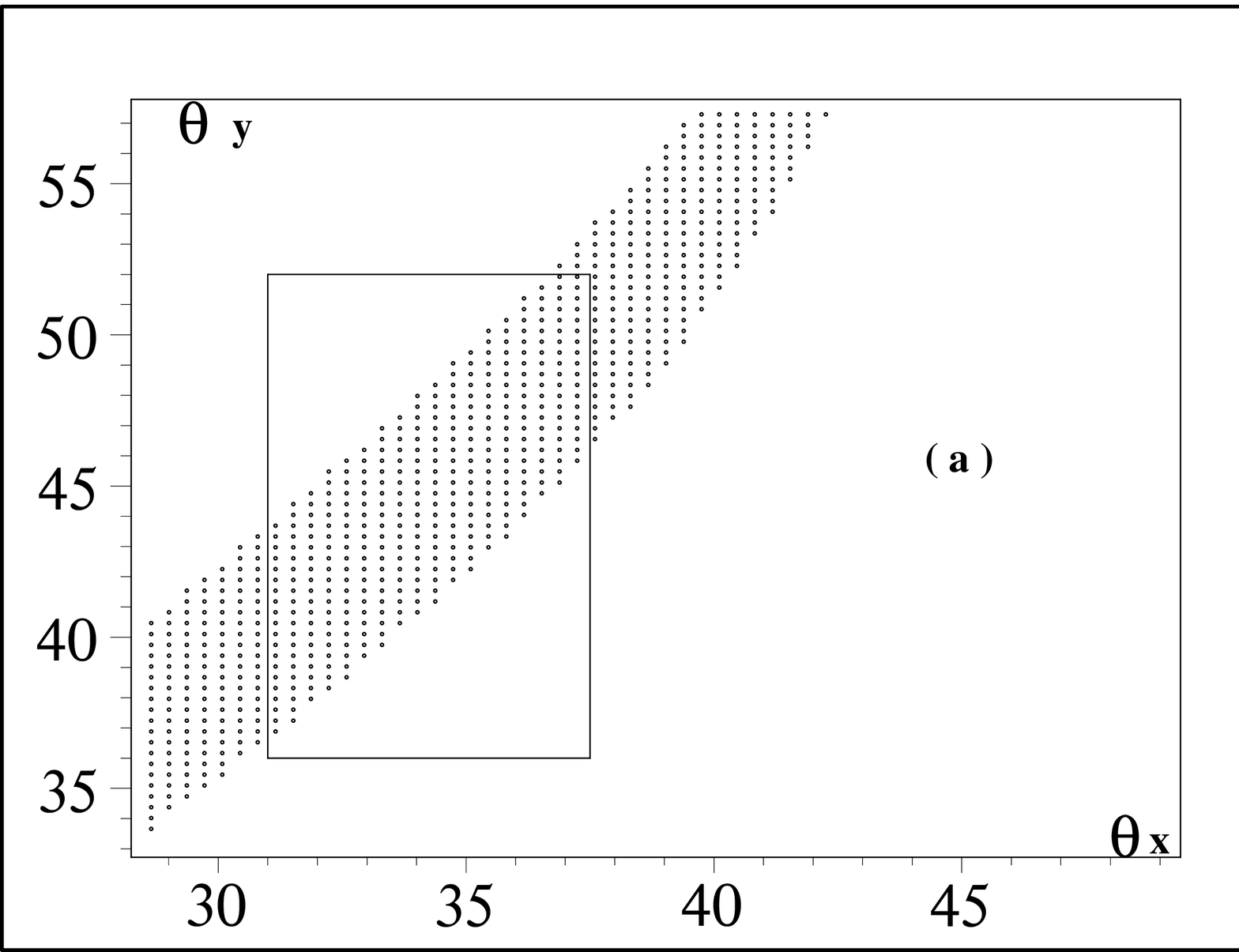}}
\end{minipage}%
\begin{minipage}[c]{0.5\textwidth}
\epsfxsize=5.5cm
\centerline{\epsfbox{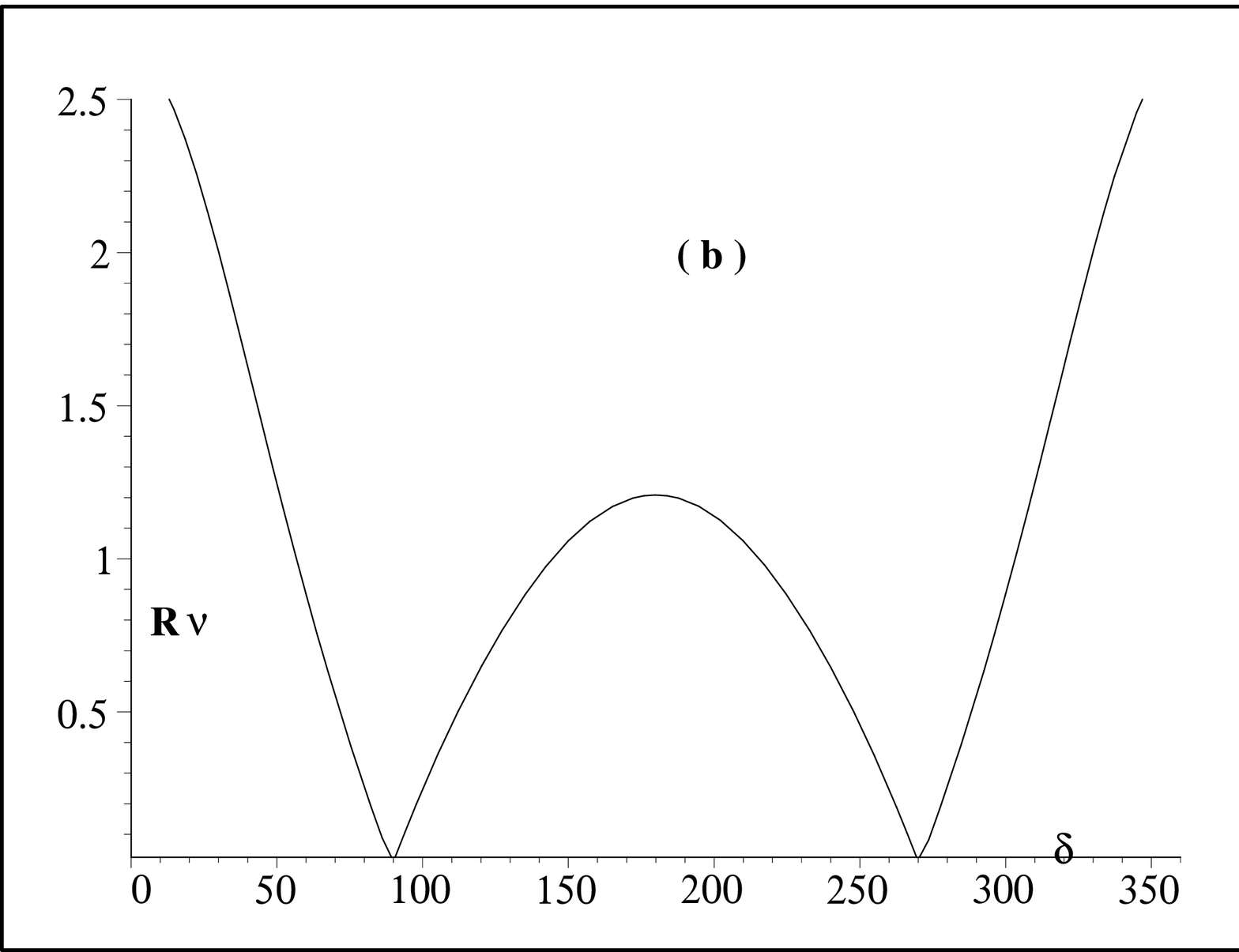}}
\end{minipage}
\vspace{0.5cm}
\caption{{\footnotesize a: The available $\theta_x,\theta_y$ parameter space for the
case
{\bf A1} pattern with $\d=0^\circ, \t_z = 7^\circ$. Both angles $\t_x$ and $\t_y$ are spanned from $29^0$ to
$57^0$.The dots represent the points satisfying all the mass-related constraints, whereas the rectangle delimit
the acceptable region for $\t_x$ and $\t_y$. The dots inside the rectangle are the acceptable points. b: $R_\nu$ as a function of
$\delta$, for $\theta_x = 34^\circ , \theta_y = 42^\circ ,
\theta_z = 5^\circ$, for case {\bf B3}, the values around $\delta=90^\circ , 270^\circ$ are
singled out. }}
\label{figA1}
\end{figure}

Numerically, we find that the estimated mass matrix has a
structure of a two-zero texture, as mentioned above, and that we have here a strong
hierarchy $m_2 \ll m_3$.
%In fact the formulae
%(\ref{xingA1-1}-\ref{xingA1-2}) are identical to those of the
%two-zero texture denoted by (A1) in \cite{Xing}.
 The pattern is
acceptable, but the numerical fitting, for ($\t_x = 34 ^\circ , \t_y = 42^\circ$), is not possible for
$\theta_z$ quite small, and only for $\th_z$ larger than $7^\circ$
we can accommodate, without tuning, the data. In figure~(\ref{figA1}.a) we show the
parameter space of ($\t_x,\t_y$) for ($\d =0^0, \t_z = 7^0$) where the rectangle delimit the experimentally acceptable region.

%%%%%%%%%%%%%%%%%%%%%%%%%%%Fig A1 %%%%%%%%%%%%%%%%%%%%%%%%%%%%
%%%%%%%%%%%%%%%%%%%%%%%%%%%%%%%%%%%%%%%%%%%%%%%%%%%%%%%%%%%%%%%%%%%%%%%%%%%%%%%%%%%%%%

One can show analytically that an acceptable phenomenology for this two-vanishing minors texture
would exclude the non-invertible mass matrix possibility and would lead automatically to a two-zero entries texture.
In fact, the conditions in eq.~(\ref{det1}) and eq.(\ref{det2}) reading now as:
\bea \label{eq1}
M_{\n\;11}\;M_{\n\;22}-M_{\n\;12}\;M_{\n\;21}&=&0\;\;\;\; (C_{33}=0),
\eea
\bea
\label{eq2}
M_{\n\;11}\;M_{\n\;23}-M_{\n\;13}\;M_{\n\;21}&=&0\;\;\;\; (C_{32}=0),
\eea
will imply, with a non-zero $M_{\n\;11}$ or $M_{\n\;12}$, the vanishing of the the minor
\bea \label{minor}C_{31}=M_{\n\;12}\;M_{\n\;23}-
M_{\n\;13}\;M_{\n\;22}\eea
which,
together with $C_{33}=C_{32}=0$, means that the mass matrix is singular.
Actually, a non-zero $M_{\n\;11}$, say, leads to either $M_{\n\;22}=0$, and hence, from eq. \ref{eq1}, to  $M_{\n\;21}=0$
implying the vanishing of $C_{31}$, or to $M_{\n\;22}\neq 0$ and so equation \ref{eq1} leads to
\bea \label{eq3} M_{\n\;11} &=& {M_{\n\;12}\; M_{\n\;12}\over M_{\n\;22}}. \eea
In the latter case ($M_{\n\;22}\neq 0$), we have two possibilities. The first is that $M_{\n\;23}=0$
which, via eq. \ref{eq2}, leads to either $M_{\n\;13}=0$ and hence $C_{31}=0$, or $M_{\n\;12}=0$ but then, from eq. \ref{eq3},
we have a vanishing $M_{\n\;11}$. The second possibility is that $M_{\n\;23}\neq0$ and hence we have (see eq. \ref{eq2})
\bea \label{eq4}M_{\n\;11}  &=& {M_{\n\;13}\; M_{\n\;21}\over M_{\n\;23}}\eea
The two equations (\ref{eq3} and \ref{eq4}) imply again either a vanishing ($M_{\n\;21}$ and $M_{\n\;11}$) or
the minor
$C_{31}$ equals to zero. We could check also that a non-zero $M_{\n\;12}$ leads again to a
vanishing-determinant matrix.
However, a consistent, phenomenologically acceptable, solution with one of the masses $m_1$ or $m_3$ equal to zero, could not be found.
Actually, the vanishing of $m_3$ in this texture leads to:
\bea \label{m30A1} K_3 &=& c_z^2 c_y^2\, e^{-2\,i\delta} =0\;\;\; (\mbox{following from}\; C_{33}) \nn \\
L_3 &=& - c_z^2 c_y s_y  \, e^{-2\,i\delta} =0\;\;\; (\mbox{following from}\; C_{32})
\eea
which is satisfied by the non-admissible values $\t_z$ or $\t_y$ equal to $\frac{\pi}{2}$.

As to the vanishing of $m_1$, it leads to  more involved expressions::
\bea \label{m10A1} K_1 &=&\left( -s_x s_y + s_z c_x c_y \, e^{-\,i\delta}\right)^2 =0       
            \;\;\; (\mbox{following from}\; C_{33}), \nn \\
L_1 &=& \left(s_x c_y  + s_z c_x s_y  \, e^{-\,i\delta}\right)
\left(s_x s_y  - s_z c_x c_y  \, e^{-\,i\delta}\right)  =0\;\;\; (\mbox{following from}\; C_{32}),
\eea
These two conditions are met if and only if $\left(s_x s_y  - s_z c_x c_y  \, e^{-\,i\delta}\right)  =0$. 
However, putting the imaginary part of the
last expression equal to zero, we see directly, since $c_x c_y \neq 0$, that $s_{\d} = 0$ otherwise
$s_z=0$ implying the phenomenologically rejected possibility $s_x s_y = 0$. Now, with the angles $\t_x,\t_y,\t_z$ being
in the first quarter, we should have $\d = 0$ leading to  
\begin{equation}
t_x = {s_z \over t_y},
\end{equation}
One can easily check that for acceptable choices of the parameters $\t_x, \t_y$ and $\t_z$ the above equation can not be satisfied.

\vspace{0.5cm}

%%%%%%%%%%%%%%%%%%%%%%%%%%%%%%%%%%%%%%%%%%%%%%%%%%%%%%%%%%%%%%%%%%%%%%%%%%%%%%
%%%%%%%%%%%%%%%%%%%%%%%%%%%%%%%%%%%%%%%%%%%%%%%%%%%%%%%%%%%%%%%%%%%%%%%%%%%%%%

%%%%%%%%%%%%%%%%%%%%%%%%%%%%%%%%%%%%%%% A2 %%%%%%%%%%%%%%%%%%%%%%
%%%%%%%%%%%%%%%%%%%%%%%%%%%%%%%%%%%%%%%%%%%%%%%%%%%%%%%%%%%%%%%%%%%%%

{\it Pattern} {\bf A2}: vanishing minors ($C_{22},C_{32}$): We get

The analytical expressions, and the representative numerical results, of the pattern {\bf A1} are valid here after the substitution dictated by eqs.
(\ref{T}, \ref{invariance}):
($c_y \leftrightarrow s_y$, $c_\d \rightarrow -c_\d$, $s_\d \rightarrow -s_\d$) and interchanging the mass matrix indices ($2 \leftrightarrow 3$) .
The conclusions stay the same, but with a different lower bound $\t_z > 4.25^\circ$. As to the parameter
space of ($\t_x,\t_y$), one should do a symmetry with respect to the line ($\t_y=45^\circ$) in figure \ref{figA1}.a
to get the parameter space in this pattern. It would be difficult to distinguish experimentally between the two patterns in class A.

%%%%%%%%%%%%%%%%%%%%%%%%%%%%%%%%%%%%%%%%%%%%%%%%%%%%%%%%%%%%%%%%%%%%%%%%%%%
%%%%%%%%%%%%%%%%%%%%%%%%%%%%%%%%%%%%%%%%%%%%%%%%%%%%%%%%%%%%%%%%%%%%%%%%%%%

\section{Class B}

%%%%%%%%%%%%%%%%%%%%%%%%%%%%%%%%%%%%%%%%%%%%%%%%%%%%%    Norm1case1  %%%%%%%%%%%%%%%%%%%%%%%%%%%%%%%%%%%%%%%%%%%%%%%%%%%
%%%%%%%%%%%%%%%%%%%%%%%%%%%%%%%%%%%%%%%%%%%%%%%%%%%%%%%%%%%%%%%%%%%%%%%%%%%%%%%%%%%%%%%%%%%%%%%%%%%%%%%%%%%%%%%%%%%%%

{\it Pattern} {\bf B3}: vanishing minors($C_{33},C_{31}$): We get
\bea
\label{xingB3-1} \frac{\lambda_1}{\lambda_3}
&=&
-\frac{s_y}{c_y}.\frac{s_xs_y-c_xc_ys_ze^{-i\d}}{s_xc_y+c_xs_ys_ze^{i\d}}e^{2i\d},
 \nonumber \\
\frac{\lambda_2}{\lambda_3} &=&
-\frac{s_y}{c_y}.\frac{c_xs_y+s_xc_ys_ze^{-i\d}}{c_xc_y-s_xs_ys_ze^{i\d}}e^{2i\d}.
 \eea

We have the following analytical approximations:
\bea
\label{xingB3-2}
\frac{m_1}{m_3} &=& t_y^2 - \frac{s_y c_\d s_z}{t_x c_y^3}+O \left( s_z^2 \right),\nonumber \\
\frac{m_2}{m_3} &=& t_y^2+  \frac{t_x s_y c_\d  s_z}{ c_y^3} +O \left( s_z^2 \right), \nonumber \\
\frac{m_2}{m_3} - \frac{m_1}{m_3} &=& \frac{s_y c_\d s_z}{c_y^3} (t_x + \frac{1}{t_x})  +O \left( s_z^2 \right), \nonumber \\
R_\nu &\approx& \left| \frac{1+t_x^2}{t_x}t^2_y t_{2y}c_\d s_z \right |
+O \left({s_z}^{2} \right),
\nonumber \\
\rho &=&\delta +\frac{s_\d s_z}{t_{2y}t_x}+ O \left( s_z \right) \left( \mbox{mod}\, \frac{\pi}{2} \right), \nonumber \\
\sigma &=& \delta -\frac{t_x s_\d s_z}{t_{2y}}+ O \left( s_z \right)\left( \mbox{mod}\, \frac{\pi}{2} \right),\nonumber \\
\r -\s &=& \frac{s_\d s_z}{t_{2y}} (t_x + \frac{1}{t_x}) + O \left( s_z \right) \left( \mbox{mod}\, \frac{\pi}{2} \right),
 \nn \\
\frac{\langle m \rangle_{ee}}{m_3} &\approx& t_y^2 + O \left( s_z \right), \nonumber \\
\frac{\langle m \rangle_{e}}{m_3} &\approx& t_y^2  + O \left(s_z\right).
\eea

In order that $m_2$ be larger than $m_1$, we see that the parameter $\d$ is restricted to be in the first and fourth quadrants.We see also
 that N-type (I-type) --hierarchy can be obtained
if $\t_y < \frac{\pi}{4}$ ($\t_y > \frac{\pi}{4}$). For $\t_z$ not too small, satisfying the $R_\n$--constraint singles out the right angles
($\frac{\pi}{2},\frac{3\pi}{2} $) for $\d$, as can be seen in figure~ (\ref{figA1}.b). When we approach the maximal mixing
limit ($\t_y = \frac{\pi}{4}$) then, from the $R_\n$ expression,  $\t_z$ tends to zero or $\d$ tends to a right angle.
However, from eq. \ref{xingB3-1}, the limit is not attained
since ($\t_y = \frac{\pi}{4}$) leads to a
degenerate spectrum ($m_1 = m_2 = m_3$). No lower bounds on $\t_z$ can be obtained.

\begin{figure}[hbtp]
\centering
\begin{minipage}[c]{0.5\textwidth}
\epsfxsize=5.5cm
\centerline{\epsfbox{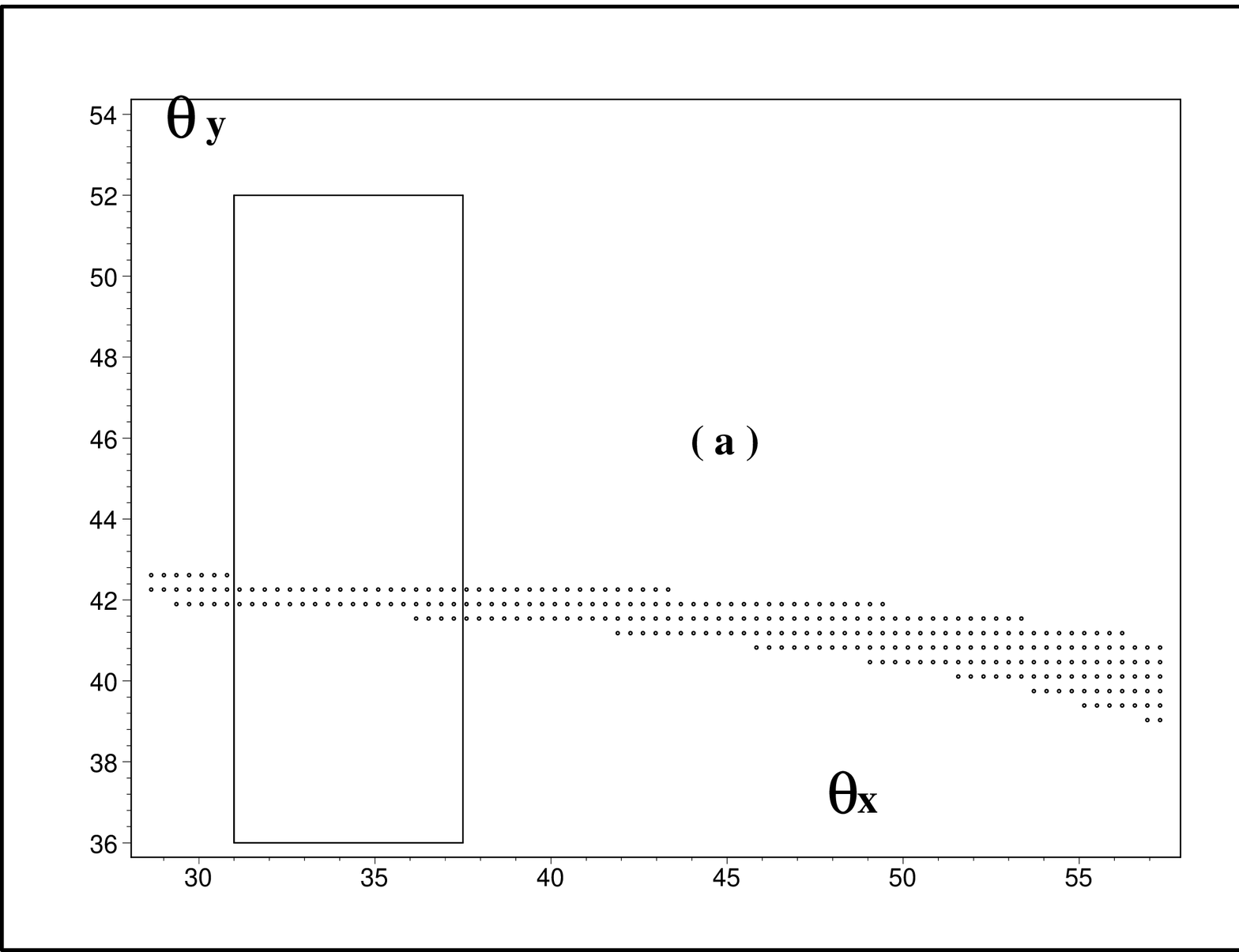}}
\end{minipage}%
\begin{minipage}[c]{0.5\textwidth}
\epsfxsize=5.5cm
\centerline{\epsfbox{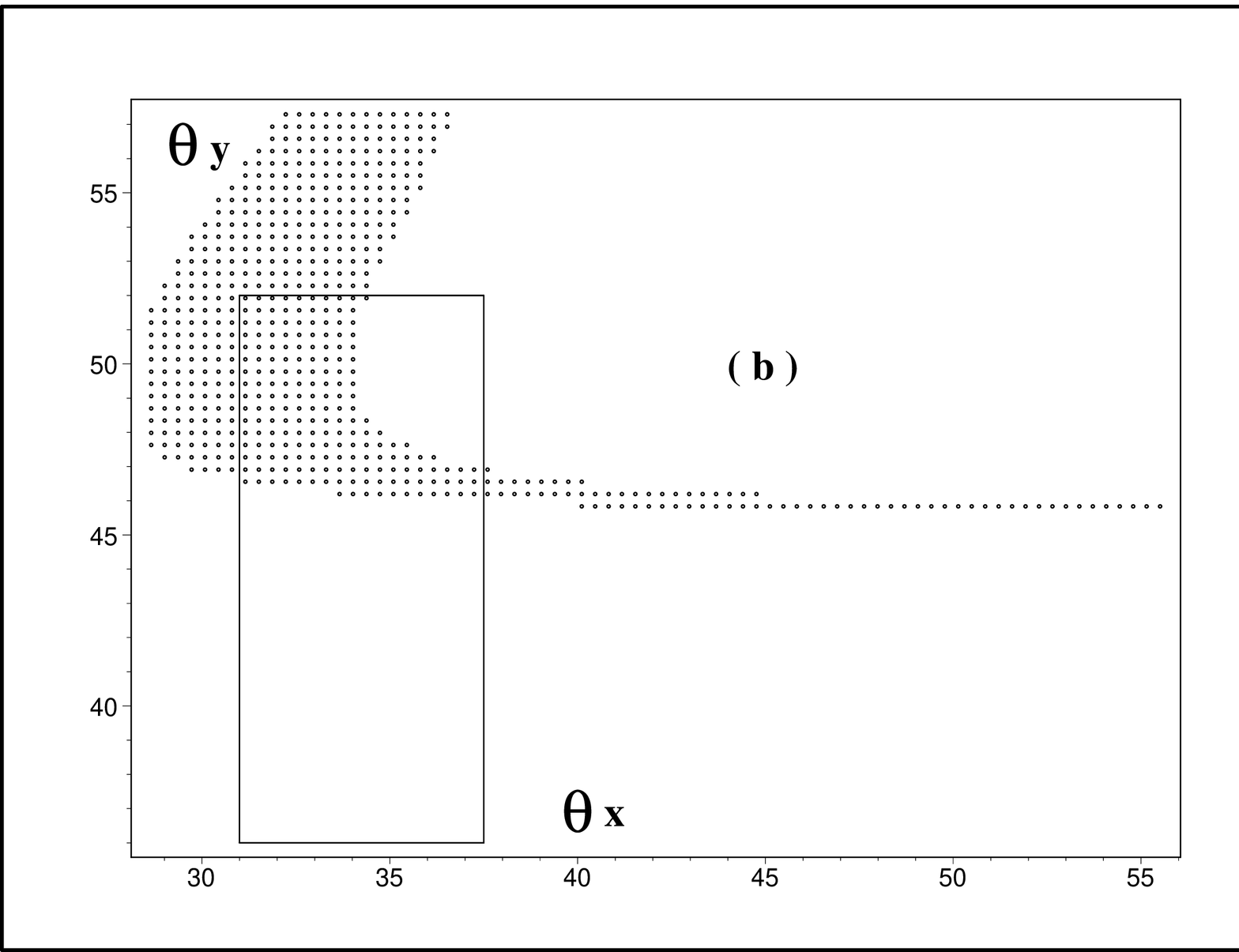}}
\end{minipage}
\vspace{0.5cm}
\caption{{\footnotesize The available $\theta_x,\theta_y$ parameter space for the
{\bf B3} pattern (a) N-type hierarchy with $\d=88.3^\circ$ and (b) I-type hierarchy with $\d=270.4^\circ$.
The range from $29^0$ to
$57^0$ is spanned for both angles $\t_x$ and $\t_y$.
The dots represent the set of points
$(\t_x, \t_y)$ which satisfy all the mass-related constraints. The rectangle is the acceptable region for $\t_x, \t_y$.
The points inside the rectangle are the acceptable points.  }}
\label{figB3}
\end{figure}

For the N-type, we take the representative point $(\t_x = 34^0, \t_y
= 42^0, \t_z = 5^0)$, and find that the $R_\n$ condition constrains the Dirac phase to be  around $\d = 88.3^\circ$. For these inputs we obtain
$m_1/m_3 = 0.8073998611$, $m_2/m_3 = 0.8141569633$, $\r  =
179.093^0$, $\s = 177.953^0$ and $R_\nu = 0.0325$. The mass $m_3$
fitted from the observed $\D m^2_{\mbox{sol}}$ is then $m_3 =
0.085\; \mbox{eV}$, whereas the values for the other remaining
parameters would be $\D m^2_{\mbox{atm}}=2.5\times
10^{-3}\;\mbox{eV}^2$, $\langle m \rangle_e =0.069\;\mbox{eV}$, $\langle m \rangle_{ee} =0.069
\;\mbox{eV}$ and $\Sigma =0.223\;\mbox{eV} $.  The numerically estimated mass matrix $M$ is \be M_\nu= m_3\left(
\begin {array}{ccc} 0.810-0.0355\,i& -0.213\times 10^{-8}+0.125\times 10^{-7}\,i&
0.022+0.417\times 10^{-2}\,i
\\\noalign{\medskip}
-0.213\times 10^{-8}+0.125\times 10^{-7}\,i&-0.99\times
10^{-8}-0.202\times 10^{-8}\,i& 0.900+0.334\times 10^{-9}\,i
\\\noalign{\medskip}
 0.022+0.417\times 10^{-2}\,i&0.900+0.334\times 10^{-9}\,i& 0.187-0.491\times 10^{-3}\,i
\end {array}
\right)
\label{massB3nor}
\ee
For the choice $\delta = 88.3^\circ$
and $\theta_z = 5^\circ$, the acceptable region in the parameter
space of $\theta_x, \theta_y$ is depicted in figure~
(\ref{figB3}.a) which indicates that no tuning
is required for the mixing angles $\theta_x$ and $\theta_y$ to
assure their consistency with the data.

%%%%%%%%%%%%%%%%%%%%%%%%%%%Fig B3 %%%%%%%%%%%%%%%%%%%%%%%%%%%%
%%%%%%%%%%%%%%%%%%%%%%%%%%%%%%%%%%%%%%%%%%%%%%%%%%%%%%%%%%%%%%%%%%%%%%%%%%%%%%%

For the I-type hierarchy, we take the representative point $(\t_x = 34^0, \t_y
= 48^0, \t_z = 5^0)$, then the $R_\n$--constraint singles out the value $\d = 270.4^\circ$.
For these inputs we obtain $m_1/m_3 = 1.226998357$, $m_2/m_3 =
1.233576775$, $\r  = 1.16^0$, $\s = 0.05^0$ and $R_\nu = 0.031$.
The mass $m_3$ fitted from the observed $\D m^2_{\mbox{sol}}$ is $m_3 = 0.07\;
\mbox{eV}$. Then the derived values for the other remaining
parameters are $\D m^2_{\mbox{atm}}=2.6\times 10^{-3}\;\mbox{eV}^2$, $\langle m \rangle_e
=0.086\;\mbox{eV}$, $\langle m \rangle_{ee} =0.086\;\mbox{eV}$ and $\Sigma =0.242\;\mbox{eV} $.
In this I-type case, the numerically
estimated mass matrix $M$ is
\be
M_\nu=
m_3\,\left( \begin {array}{ccc}  1.227+ 0.034\,i&- 0.35\times 10^{-9}+ 0.6\times 10^{-9}\,i&- 0.030- 0.005\,i
\\\noalign{\medskip}- 0.35\times 10^{-9}+ 0.6\times 10^{-9}\,i&- 0.33\times 10^{-8}-{
 6.0\times 10^{-11}}\,i& 1.111+{ 2.247\times 10^{-12}}\,i
\\\noalign{\medskip}- 0.030- 0.005\,i& 1.111 +{
 2.247\times 10^{-12}}\,i&- 0.226+ 0.001\,i
\end {array} \right)
\label{massB3inv}
\ee
The parameter space of $\theta_x, \theta_y$, for the choice $\delta = 270.4^\circ$
and $\theta_z = 5^\circ$, is depicted in figure~
(\ref{figB3}.b). We see that it is quite large and no tuning is necessary.

The numerical solutions in both types of hierarchy reproduce a two-zero texture, which can
accommodate acceptably the data. As in class {A}, one could show analytically that a two-vanishing minors such texture with
non-vanishing entries leads to a singular mass matrix. Nevertheless, as in class {A}, no consistent solution could be found when attaining the limit of
one vanishing mass.
\vspace{0.5cm}
%%%%%%%%%%%%%%%%%%%%%%%%%%%%%%%%%%%%%%%%%%%%%%%%%%%%%%%%%%%%%%%%%%%%%%%%%%%%%%%
%%%%%%%%%%%%%%%%%%%%%%%%%%%%%%%%%%%%%%%%%%%%%%%%%%%%%%%%%%%%%%%%%%%%%%%%%%%%%%%

%%%%%%%%%%%%%%%%%%%%%%%%%%%%%%%%%%%%%%%%%%%%%%%%%%%%%%%%%%%%%%%%%%%%%%%%%%%%%%%%
%%%%%%%%%%%%%%%%%%%%%%%%%%%%%%%%  B4 %%%%%%%%%%%%%%%%%%%%%%%%%%%%%%%%%%%%%%%%%%
{\it Pattern} {\bf B4}: vanishing minors ($C_{22},C_{21}$):

The analytical expressions, and the representative numerical results, of the pattern {\bf B3} are valid here after the substitution
prescribed by eqs.
(\ref{T}, \ref{invariance}).
The conclusions are similar in that $\d$ is now restricted to the second and third quadrants,
and that we have N-type (I-type) hierarchy when $\t_y$ is larger (smaller) than $\frac{\pi}{4}$.
The parameter
space of ($\t_x,\t_y$), depicted in figure ( \ref{figB3}.b), should
be reflected through the line ($\t_y=45^\circ$) to get what corresponds {\bf B4}--pattern.
\vspace{0.5cm}
%%%%%%%%%%%%%%%%%%%%%%%%%%%%%%%%%%%%%%%%%%%%%%%%%%%%%%%%%%%%%%%%%%%%%%%%%%%%%%%%%%%
%%%%%%%%%%%%%%%%%%%%%%%%%%%%%%%%%%%%%%%%%%%%%%%%%%%%%%%%%%%%%%%%%%%%%%%%%%%%%%%%%%%

%%%%%%%%%%%%%%%%%%%%%%%%%%%%%%%%%%%%%%%%%%%%%%%%%%%%   case B5  %%%%%%%%%%%%%%%%%%%%%%%%%%%%%%%%%%%%%%%%%%%%%%%%%%%
%%%%%%%%%%%%%%%%%%%%%%%%%%%%%%%%%%%%%%%%%%%%%%%%%%%%%%%%%%%%%%%%%%%%%%%%%%%%%%%%%%%%%%%%%%%%%%%%%%%%%%%%%%%%%%%%%%%%%

{\it Pattern} {\bf B5}: vanishing minors ($C_{33},C_{12}$): We get \bea
\label{lam1n2} \frac{\lambda_1}{\lambda_3} &=&
\frac{s_x\,c_x\,c_y\,s_z^2\,\left(1+s_y^2\right)\,e^{-2\,i\,\delta}
+s_z\,s_y^3\,c_{2x}\,e^{-\,i\,\delta} + s_x\,c_y\,c_x\,s_y^2}
{s_z\,s_y\,c_y^2\,s_x^2\,e^{-3\,i\,\delta} +\left[-
c_x\,c_y^3\,s_{x}\,\left(1+s_z^2\right) +
2\,s_x\,c_y\,c_x\,s_z^2\right]\,e^{-2\,i\,\delta} +
\,s_z\,c_x^2\,s_y^3\,e^{-i\,\d}}
 ,\nonumber \\
\frac{\lambda_2}{\lambda_3} &=&
\frac{-s_x\,c_y\,c_x\,s_z^2\,\left(1+s_y^2\right)\,e^{-2\,i\,\delta}
-s_y^3\,s_z\,c_{2x}\,e^{-i\,\delta}-c_x\,s_x\,c_y\,s_y^2}{s_z\,s_y\,c_y^2\,c_x^2\,
e^{-3\,i\,\d}+\left[-2\,s_x\,c_x\,c_y\,s_z^2 + s_x\,c_x\,c_y^3\,
\left(1+s_z^2\right) \right]\,e^{-2\,i\,\d} +
s_z\,s_x^2\,s_y^3\,e^{-i\,\d}}.
 \eea
We have the following analytical approximations:
\bea
\frac{m_1}{m_3} &=& t_y^2 + \frac{t_y^3 c_\d s_z}{c_y^2 t_x}+O \left( s_z^2 \right), \nonumber \\
\frac{m_2}{m_3} &=& t_y^2 -\frac{t_y^3 c_\d t_x s_z}{c_y^2} +O \left( s_z^2 \right),\nonumber \\
\frac{m_2}{m_3} - \frac{m_1}{m_3} &=& -\frac{c_\d t_y^3 s_z}{c_y^2} (t_x + \frac{1}{t_x}) ++O \left( s_z^2\right),\nonumber \\
R_\nu &=& \left | \frac{-4 s_y^2 t_y^3 c_\d}{c_{2x} c_{2y}}  s_z  \right |+O \left( s_z^2\right),\nonumber \\
\rho &=& \delta -\frac{t_x c_{2y} s_\d}{2 c_y^2 t_y} s_z + O \left( s_z^2 \right) \left( \mbox{mod}\, \frac{\pi}{2} \right),\nonumber \\
\sigma &=& \delta +\frac{c_{2y} t_y s_\d}{2 c_y^2 t_x} s_z + O \left( s_z^2 \right) \left( \mbox{mod}\, \frac{\pi}{2} \right),\nonumber \\
\r -\s &=&  -\frac{c_{2y}s_\d}{2 c_y^2} (\frac{t_x}{t_y}+\frac{t_y}{t_x}) s_z + O \left( s_z^2 \right) \left( \mbox{mod}\, \frac{\pi}{2} \right),\nonumber \\
\frac{\langle m \rangle_{ee}}{m_3} &=& t_y^2 +O \left( s_z
\right),\nonumber \\
\frac{\langle m \rangle_{e}}{m_3} &=& t_y^2 +O \left( s_z
\right). \eea

This pattern is not a two-texture zero \cite{Lavoura}, and the parameter $\d$ is restricted to be in the second and third quadrants.
The N-type (I-type) --hierarchy can be obtained
if $\t_y < \frac{\pi}{4}$ ($\t_y > \frac{\pi}{4}$). We see that, for $\t_z$ not too small, the angle $\d$ needs to be
peaked around a right angle in order to  satisfy the $R_\n$--constraint. The maximal mixing limit can not be reached since
($\t_y = \frac{\pi}{4}$) leads to a
degenerate spectrum ($m_1 = m_2 = m_3$). However, when ($\t_y \rightarrow \frac{\pi}{4}$) then $\t_z$ tends to zero or $\d$ tends to
a right angle. No lower bounds on $\t_z$ can be obtained.

\begin{figure}[hbtp]
\centering
\begin{minipage}[c]{0.5\textwidth}
\epsfxsize=5.5cm
\centerline{\epsfbox{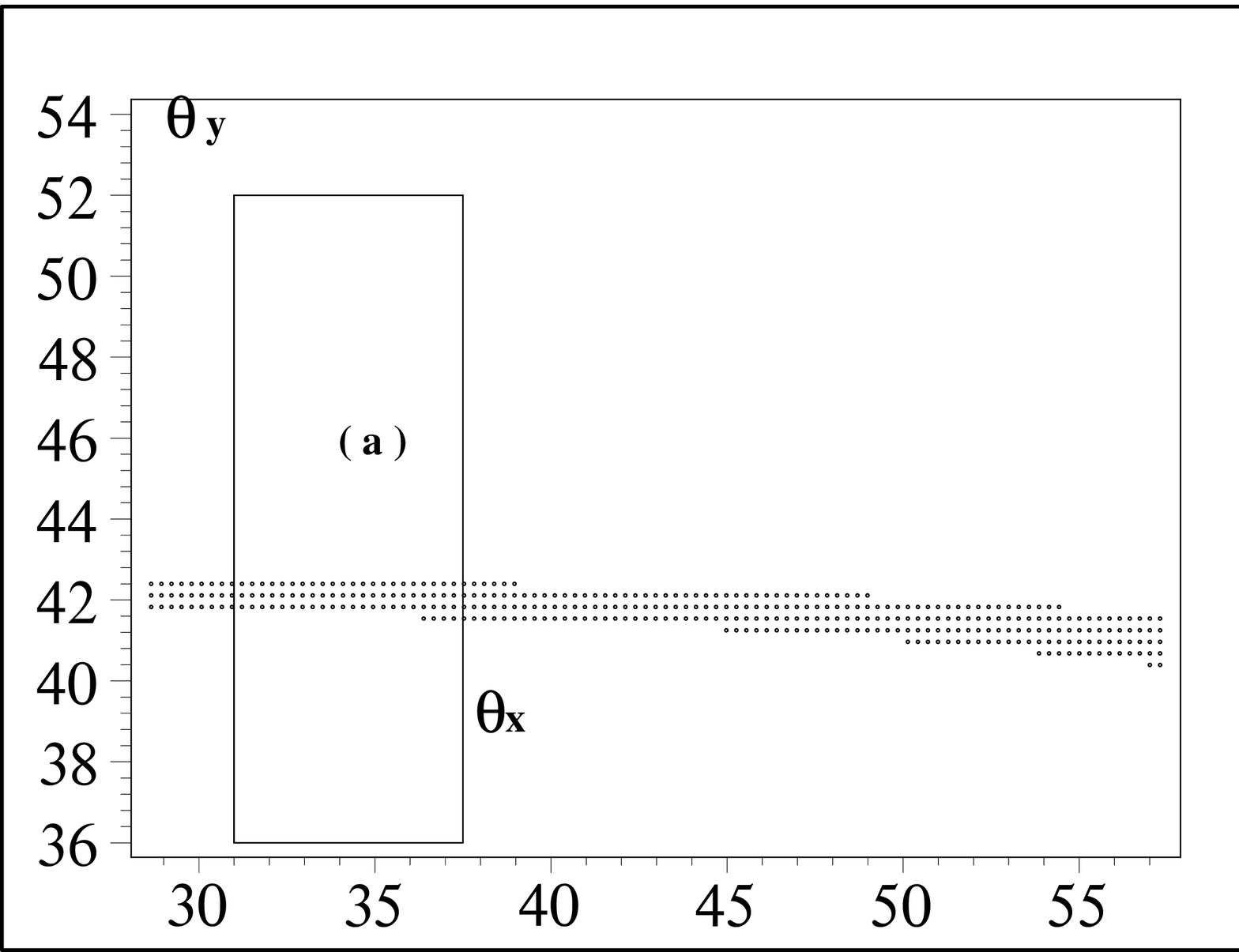}}
\end{minipage}%
\begin{minipage}[c]{0.5\textwidth}
\epsfxsize=5.5cm
\centerline{\epsfbox{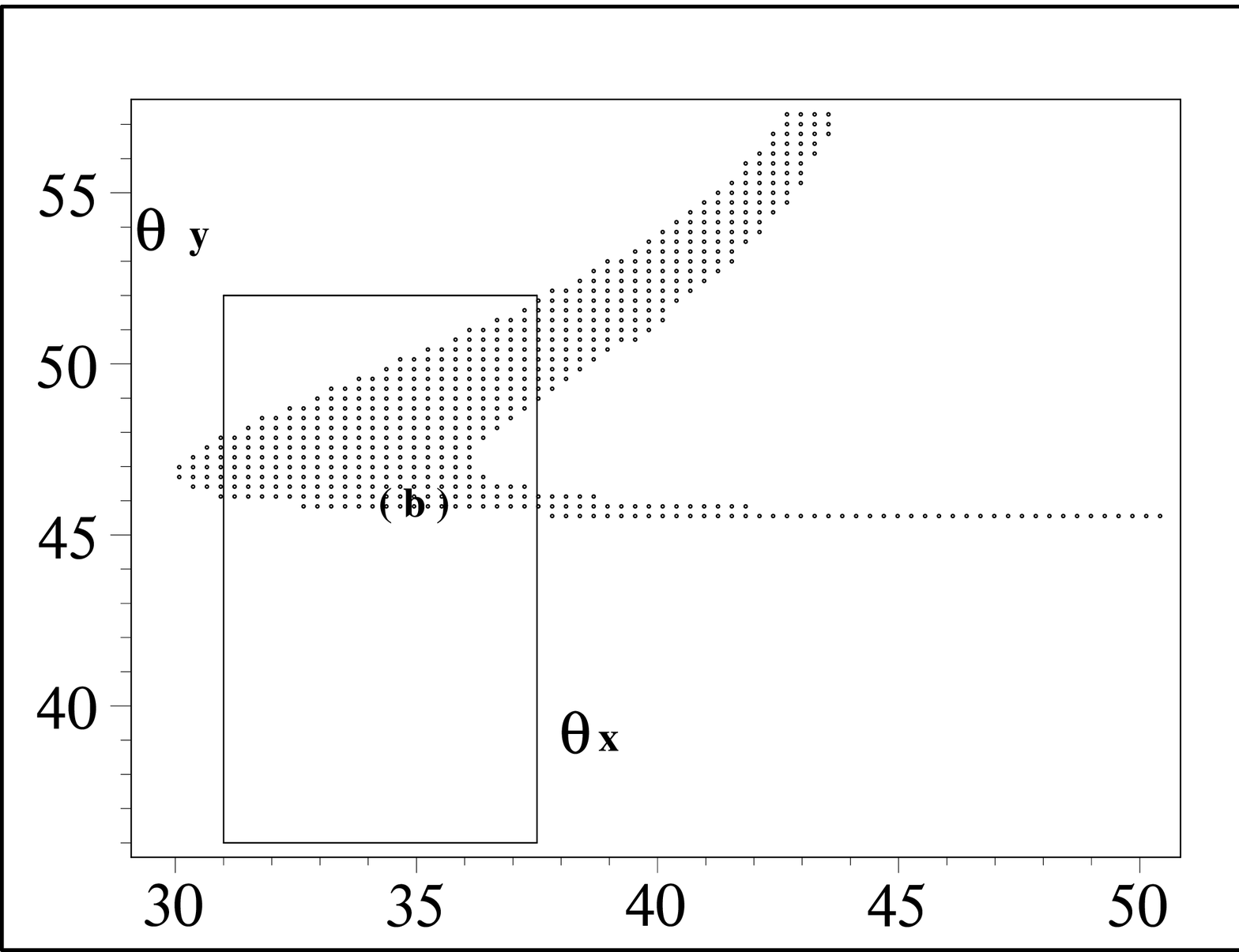}}
\end{minipage}
\vspace{0.5cm}
\caption{{\footnotesize The available $\theta_x,\theta_y$ parameter space for the
{\bf B5} pattern (a) N-type hierarchy with $\d=92^\circ$ and (b) I-type hierarchy with $\d=269.83^\circ$.
The range from $29^0$ to
$57^0$ is spanned for both angles $\t_x$ and $\t_y$.
The dots represent the set of points
$(\t_x, \t_y)$ which satisfy all the mass-related constraints. The rectangle is the acceptable region for $\t_x, \t_y$.
The points inside the rectangle are the acceptable points.  }}
\label{figB5}
\end{figure}

For the N-type, we take the representative point $(\t_x = 34^0, \t_y =
42^0, \t_z = 5^0)$, and the the $R_\n$--condition constrains $\d$ to be around $92^\circ$. For these inputs we obtain $m_1/m_3
= 0.8018004128$, $m_2/m_3 = 0.8090251390$, $\r  = 1.4^0$, $\s =
2.34^0$ and $R_\nu = 0.0336$. The mass $m_3$ fitted from the
observed $\D m^2_{\mbox{sol}}$ is $m_3 = 0.082\; \mbox{eV}$, and
then the values for the other remaining parameters are inferred to
be $\D m^2_{\mbox{atm}}=2.4\times 10^{-3}\;\mbox{eV}^2$, $\langle m \rangle_e
=0.066\;\mbox{eV}$, $\langle m \rangle_{ee} =0.066\;\mbox{eV}$ and $\Sigma
=0.215\;\mbox{eV} $. The numerically estimated mass matrix $M$, in
this pattern, is \be M_\nu= m_3\,\left(
\begin {array}{ccc}
0.804+0.047\,i&0.021-0.495\times 10^{-2}\,i
& 0.441\times 10^{-2}-0.107\times 10^{-2}\,i\\
0.021-0.495\times 10^{-2}\,i& 0.487\times 10^{-3}-0.284\times 10^{-3}\,i
& 0.898+0.839\times 10^{-3}\,i\\
 0.441\times 10^{-2}-0.107\times 10^{-2}&0.898+0.839\times 10^{-3}\,i
 & 0.191-0.629\times 10^{-3}\,i
 \end {array}
\right).
\label{massB5nor}
\ee
For the choice $\delta = 92^\circ$
and $\theta_z = 5^\circ$, the acceptable region in the parameter
space of $\theta_x, \theta_y$ is quite large and is depicted in figure~
(\ref{figB5}.a), thus no tuning is necessary for the mixing angles $\theta_x$ and $\theta_y$ to
assure their consistency with the data.

%%%%%%%%%%%%%%%%%%%%%%%%%%%Fig B5 %%%%%%%%%%%%%%%%%%%%%%%%%%%%
%%%%%%%%%%%%%%%%%%%%%%%%%%%%%%%%%%%%%%%%%%%%%%%%%%%%%%%%%%%%%%%%%%%%%%%%%%%%%%%

For the I-type hierarchy, we take the representative point $(\t_x = 34^0, \t_y
= 48^0, \t_z = 5^0)$, then the $R_\n$--constraint singles out the value $\d = 269.83^\circ$.
For these inputs we obtain $m_1/m_3 = 1.234417996$, $m_2/m_3 =
1.241674367$, $\r  = 178.84^\circ$, $\s = 0.26^\circ$ and $R_\nu = 0.033$.
The mass $m_3$ fitted from the observed $\D m^2_{\mbox{sol}}$ is $m_3 = 0.066\;
\mbox{eV}$. Then the derived values for the other remaining
parameters are $\D m^2_{\mbox{atm}}=2.4\times 10^{-3}\;\mbox{eV}^2$, $\langle m \rangle_e
=0.082\;\mbox{eV}$, $\langle m \rangle_{ee} =0.082\;\mbox{eV}$ and $\Sigma =0.231\;\mbox{eV} $.
In this I-type case, the numerically
estimated mass matrix $M$ is
\be
M_\nu=
m_3\,\left( \begin {array}{ccc}  1.234- 0.03055\,i&- 0.03414+ 0.004469\,i& 0.007299- 0.0009687\,i\\\noalign{\medskip}- 0.03414+ 0.004469\,i&
 0.0009338- 0.0002241\,i& 1.114- 0.0003354\,i\\\noalign{\medskip}
 0.007299- 0.0009687\,i& 1.114- 0.0003354\,i&- 0.2382+ 0.0004991\,i
\end {array} \right)
\label{massB5inv}
\ee
The parameter space of $\theta_x, \theta_y$, for the choice $\delta = 269.83^\circ$
and $\theta_z = 5^\circ$, is depicted in figure~
(\ref{figB5}.b). Again it is quite large and no tuning is necessary.

The numerical solutions in both types of hierarchy are not two-zero textures. The pattern is acceptable.
However, it does not allow for non-invertible mass matrices as no consistent solution could be found in
 the limit of
one vanishing mass.
\vspace{0.5cm}
%%%%%%%%%%%%%%%%%%%%%%%%%%%%%%%%%%%%%%%%%%%%%%%%%%%%%%%%%%%%%%%%%%%%%%%%%%%%%%%%%%%%
%%%%%%%%%%%%%%%%%%%%%%%%%%%%%%%%%%%%%%%%%%%%%%%%%%%%%%%%%%%%%%%%%%%%%%%%%%%%%%%%%%%%%

%%%%%%%%%%%%%%%%%%%%%%%%%%%%%%%%%%%%%%%%%%%%%%%%%%%%%%%%%%%%%%%%%%%%%%%%%%%%%%%%
%%%%%%%%%%%%%%%%%%%%%%%%%%%%%%%%  B6 %%%%%%%%%%%%%%%%%%%%%%%%%%%%%%%%%%%%%%%%%%
{\it Pattern} {\bf B6}: vanishing minors ($C_{22},C_{31}$):

The analytical expressions, and the representative numerical results, of the pattern {\bf B5} are valid here after the substitution
imposed by eqs.
(\ref{T}, \ref{invariance}).
The conclusions are similar in that $\d$ is now restricted to the first and fourth quadrants,
and that we have N-type (I-type) hierarchy when $\t_y$ is larger (smaller) than $\frac{\pi}{4}$.
The parameter
space of ($\t_x,\t_y$), depicted in figure (\ref{figB5}.b), should be reflected through the line ($\t_y=45^\circ$) to
get what corresponds {\bf B6}--pattern.
\vspace{0.5cm}

%%%%%%%%%%%%%%%%%%%%%%%%%%%%%%%%%%%%%%%%%%%%%%%%%%%%%%%%%%%%%%%%%%%%%%%%%%%%%%%%%%%
%%%%%%%%%%%%%%%%%%%%%%%%%%%%%%%%%%%%%%%%%%%%%%%%%%%%%%%%%%%%%%%%%%%%%%%%%%%%%%%%%%%
 The class ({\bf B}) can be distinguished easily from class ({\bf A}) in that it allows for I-type hierarchy.  We summarize in
 table (\ref{classB}) the `experimental' signatures which enable to distinguish between its patterns.

\begin{table}[htbp]
\begin{center}
\begin{tabular}{|c||c|c|c|}
\hline
   $\mbox{Model}$ & $\d$--Quadrant & I-type & N-type\\
\hline
\hline
{\bf B3} & $1,4$  & $\t_y > \frac{\pi}{4}$ & $\t_y < \frac{\pi}{4}$ \\
\hline
{\bf B4} & $2,3$  & $\t_y < \frac{\pi}{4}$ & $\t_y > \frac{\pi}{4}$ \\
\hline
{\bf B5} & $2,3$  & $\t_y > \frac{\pi}{4}$ & $\t_y < \frac{\pi}{4}$ \\
\hline
{\bf B6} & $1,4$  & $\t_y < \frac{\pi}{4}$ & $\t_y > \frac{\pi}{4}$ \\
\hline
\end{tabular}
\end{center}
 \caption{\small  `Experimental' Signatures distinguishing the different {\bf B}--class patterns.}
\label{classB}
\end{table}

\vspace{0.5cm}
%%%%%%%%%%%%%%%%%%%%%%%%%%%%%%%%%%%%%%%%%%%%%%%%%%%%%%%%%%%%%%%%%%%%%%%%%%%%%%%%%%%%%%
%%%%%%%%%%%%%%%%%%%%%%%%%%%%%%%%%%%%%%%%%%%%%%%%%%%%%%%%%%%%%%%%%%%%%%%%%%%%%%%%%%%%%

\section{Class D}

%%%%%%%%%%%%%%%%%%%%%%%%%%%%%%%%% D %%%%%%%%%%%%%%%%%%%%%%%%%%%%%%%%
%%%%%%%%%%%%%%%%%%%%%%%%%%%%%%%%%%%%%%%%%%%%%%%%%%%%%%%%%%%%%%%%%%%%%%%%%%%%%

{\it Pattern} {\bf D}: vanishing minors ($C_{33},C_{22}$): We get
\bea
\frac{\lambda_1}{\lambda_3} &=& \frac{-2\,c_x\,s_x\,s_y\,c_y\,s_z^3\,e^{-2\,i\,\d}
+\left(-s_z^2\,c_{2y}+2\,c_x^2\,s_z^2\,c_{2y}\right)\,e^{-i\,\d}
 -2\,s_x\,c_x\,s_y\,c_y\,s_z}{c_z^2\,c_x\,\left(-c_x\,c_{2y} + 2\,s_x\,c_y\,s_z\,s_y\,
 e^{-i\,\d}\right)\,e^{-i\,\d}}, \nonumber\\
\frac{\lambda_2}{\lambda_3} &=& \frac{-2\,c_x\,s_x\,s_y\,c_y\,s_z^3\,e^{-2\,i\,\d}
+\left(-s_z^2\,c_{2y}+2\,c_x^2\,s_z^2\,c_{2y}\right)\,e^{-i\,\d}
 -2\,s_x\,c_x\,s_y\,c_y\,s_z}{c_z^2\,\left(s_x^2\,c_{2y} +2\,s_x\,c_y\,s_z\,
 c_x\,s_y\,e^{-i\,\d}\right)\,e^{-i\,\d}}.\nonumber \\
\eea

We have the following analytical approximations:
\bea
\frac{m_1}{m_3} &=&  t_x \left |  t_{2y} \right | s_z + O\left( s_z^2 \right),\nonumber \\
\frac{m_2}{m_3} &=&    \frac{\left |t_{2y}\right |}{t_x}  s_z + O\left( s_z^2 \right),\nonumber \\
\frac{m_2}{m_3} - \frac{m_1}{m_3} &=& \left |t_{2y}\right | (\frac{1}{t_x} - t_x) s_z + O\left( s_z^2 \right),\nonumber \\
R_\nu &=& \frac {4\,t_{2y}^2\,s_z^2}{s_{2x}\,t_{2x}}
+O \left( s_z^3\right) \approx \frac {4\,t_{x}^2 (\frac{m_2}{m_3})^2}{s_{2x}\,t_{2x}}
+O \left( s_z^3\right),\nonumber \\
\rho &=& \frac{\delta}{2}+ 2\frac{c_{2x} -c_x^2 s_{2y}^2}{s_{2x} s_{4y}} s_\d s_z + O \left( s_z^2
 \right) \left( \mbox{mod}\, \frac{\pi}{2} \right),\nonumber \\
 \sigma &=& \frac{\delta}{2}+ 2\frac{c_{2x} +s_x^2 s_{2y}^2}{s_{2x} s_{4y}} s_\d s_z +O \left( s_z^2
 \right) \left( \mbox{mod}\, \frac{\pi}{2} \right),\nonumber \\
 \r - \s &=& - \frac{t_{2y}}{s_{2x}} s_\d s_z +O \left( s_z^2
 \right) \left( \mbox{mod}\, \frac{\pi}{2} \right),\nonumber \\
\frac{\langle m \rangle_{ee}}{m_3} &=& \frac{1}{c_{2y}^2}s_z^2+O\left(s_z^3\right), \nonumber \\
\frac{\langle m \rangle_{e}}{m_3} &=&
\frac{1}{c_{2y}}s_z+O\left(s_z^2\right).
\eea
We see that one can not obtain an I-type hierarchy and satisfy the $R_\n$--constraint simultaneously.
Although the first term in the expansion of $R_\n$ in powers of $s_z$ is independent of $\d$, we find numerically that satisfying the
$R_\n$--constraint, for fixed ($\t_x, \t_y$ and $\t_z$), singles out specific values of $\d$. In fact, upon closer examination,
we found that the next term in the expansion to be of the same order of the first term, and there is a delicate cancelation enforcing the
small value of $R_\n$:
\bea \label{delicate}
R_\nu &=& \left |\frac {4\,t_{2y}^2\,s_z^2}{s_{2x}\,t_{2x}} - \frac{2c_\d t_{2y} s_z^3}{s_x^3 c_x^3 c_{2y}^2}
\left[ 1-s_{2x}^2 (1-\frac{1}{4} s_{2y}^2)\right] \right |
+O \left( s_z^4\right), \eea
and so for plausible values of $\t_x$ and $\t_y$ the expansion parameter $s_z$ might not be the right one.
Moreover, eq.(\ref{delicate})
tells us that although $\d$  is not constrained, however it is correlated to the angle $\t_y$ in that when the latter is smaller than $\frac{\pi}{4}$
the angle $\d$ needs to be in the first or fourth quadrants ($c_\d > 0$), whereas $\t_y$ being larger than $\frac{\pi}{4}$ makes the angle $\d$
to be in the second and third quadrants ($c_\d < 0$). Also, there is no lower bounds on $s_z$, however, for $s_z$ small enough the angle $\t_y$
approaches the maximal mixing limit.

%and in fact we have checked that $s_z =0$ is an acceptable choice provided $\t_y$ is fine-tuned to be too close
%to the
%maximal mixing value. Also, this value ($\t_y = \frac{\pi}{4}$) could be attained with $\t_z = 5^\circ$ and $\d$ peaked
%at $\frac{\pi}{2}$.

\begin{figure}[htpb]
\centering \epsfxsize=6cm
\centerline{\epsfbox{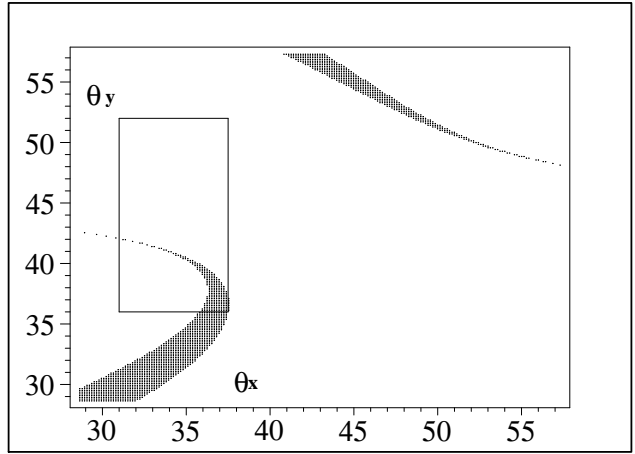}}
\vspace{0.5cm}
\caption{{\footnotesize The available $\theta_x,\theta_y$ parameter space for the
{\bf D} pattern with $\d=52.7^\circ$.  The hierarchy is of N-type, and we have spanned both $\t_x$ and $\t_y$ from $29^0$ to
$57^0$ .
The dots represent the set of points
$(\t_x, \t_y)$ which satisfy all the mass-related constraints, whereas the rectangle is the acceptable region for $\t_x, \t_y$.
The points inside the rectangle are the acceptable points.  }}
\label{figD}
\end{figure}

As a representative point for this pattern we take $(\t_x = 34^0, \t_y = 41^0, \t_z = 5^0)$, and find that the $R_\n$--constraint singles out the
value $\d = 52.7^\circ$.
For these inputs we obtain $m_1/m_3 = 0.5115705296$, $m_2/m_3 =
0.5342013580$, $\r  = 14.35^\circ$, $\s = 128.95^\circ$ and $R_\nu = 0.033$.
The mass $m_3$ fitted from the observed $\D m^2_{\mbox{sol}}$ is $m_3 = 0.0578\;
\mbox{eV}$. Then the derived values for the other remaining
parameters are $\D m^2_{\mbox{atm}}=2.40 \times 10^{-3}\;\mbox{eV}^2$, $\langle m \rangle_e
=0.030\;\mbox{eV}$, $\langle m \rangle_{ee} =0.016\;\mbox{eV}$ and $\Sigma =0.118 \;\mbox{eV} $.
In this pattern the numerically
estimated mass matrix $M$ is
\be
M_\nu=
m_3\,\left( \begin {array}{ccc}  0.279+ 0.006\,i&- 0.290- 0.007\,i& 0.336+ 0.006\,i
\\\noalign{\medskip}- 0.290- 0.007\,i& 0.302+
 0.009\,i& 0.641- 0.007\,i\\\noalign{\medskip}
 0.336+ 0.006\,i& 0.641- 0.007\,i&
 0.404+ 0.005\,i\end {array} \right)
. \label{massD} \ee
%highly constrained
The parameter space of $\theta_x, \theta_y$, for the choice $\delta = 59.3^\circ$
and $\theta_z = 5^\circ$, is depicted in figure~
(\ref{figD}). We found no consistent non-invertible such texture which can accommodate the data.

%%%%%%%%%%%%%%%%%%%%%%%%%%%Fig D %%%%%%%%%%%%%%%%%%%%%%%%%%%%
%%%%%%%%%%%%%%%%%%%%%%%%%%%%%%%%%%%%%%%%%%%%%%%%%%%%%%%%%%

This class is hardly distinguishable from the class {\bf A}, except that, contrary to the latter, the parameter $\langle m \rangle_{ee}$ is not zero,
and the mass-ratio $\frac{m_2}{m_3}$ is often larger than what we get in the  {\bf A} class.

%%%%%%%%%%%%%%%%%%%%%%%%%%%%%%%%%%%%%%%%%%%%%%%%%%%%%%%%%%%%%%%%%%%%%%%%%%%%%%%%%%%%%%%%%%%%%%
%%%%%%%%%%%%%%%%%%%%%%%%%%%%%%%%%%%%%%%%%%%%%%%%%%%%%%%%%%%%%%%%%%%%%%%%%%%%%%%%%%%%%%%%%%%%%%

\section{Singular models}

%%%%%%%%%%%%%%%%%%%%%%%%%%%%%%%%%%%%%%%% S1, S2 S3 %%%%%%%%%%%%%%%%%%%%%%%%%%%%%%%%%%%%%
%%%%%%%%%%%%%%%%%%%%%%%%%%%%%%%%%%%%%%%%%%%%%%%%%%%%%%%%%%%%%%%%%%%%%%%%%%%%%%%%%%%%

{\it Patterns} {\bf S1, S2 and S3}: vanishing minors ($C_{31},C_{11}$), ($C_{21},C_{11}$) and ($C_{31},C_{21}$):
The patterns {\bf S1} and {\bf S2} (which are related by the $T$ symmetry)  lead to $m_2 < m_1$, whence they are not acceptable.
The pattern {\bf S3} lead to a degenerate non-acceptable model with ($m_1 = m_2 = m_3$).

However, all these models can allow for one, and only one, model in which the neutrino mass matrix is singular. This happens only when
$m_3 = 0$ and $\t_z = 0$. A vanishing $\theta_z$ is still consistent with
experimental data as shown in eq.~(\ref{expdata}). In this pattern
with vanishing $m_3$ we have
$m_1=\sqrt{\Delta m^2_{\mbox{atm}} - \Delta m^2_{\mbox{sol}}}$
and $m_2=\sqrt{\Delta m^2_{\mbox{atm}}}$. As
 to the parameters $\langle m \rangle_{ee}$ and $\langle m
 \rangle_{e}$, we have respectively
 \bea
\langle m \rangle_{ee}&=& \sqrt{m_1^2\,c_x^4 + m_2^2\,s_x^4 +
2\,m_1\,m_2\, c_x^2\,s_x^2\,c_{2\rho-2\sigma}},\nonumber\\
\langle m \rangle_{e}&=& \sqrt{m_1^2\,c_x^2 + m_2^2\,s_x^2}.
 \eea
The resulting mass matrix has the following elements, where the
missing ones are related by symmetry ($M_\nu=M_\nu^T$),
\bea
M_{\nu\;11} &= & \left( m_1\,c_x^2\,e^{2\,i\,\rho} +
m_2\,s_x^2\,e^{2\,i\,\sigma}\right),
\nonumber \\
M_{\nu\;12} &= & s_x\,c_x\,c_y\,e^{-i\,\d}\,
\left(-m_1\,e^{2\,i\,\rho} + m_2\,e^{2\,i\,\sigma}\right),\nonumber \\
M_{\nu\;13} &= & s_x\,c_x\,s_y\,e^{-i\,\d}\,
\left(m_1\,e^{2\,i\,\rho} - m_2\,e^{2\,i\,\sigma}\right),\nonumber \\
M_{\nu\;22} &= & c_y^2\,e^{-2\,i\,\d}\,
\left(m_1\,s_x^2\,e^{2\,i\,\rho} + m_2\,c_x^2\,e^{2\,i\,\sigma}\right),\nonumber \\
M_{\nu\;23} &= & -  c_y\,s_y\,e^{-2\,i\,\d}\,
\left(m_1\,s_x^2\,e^{2\,i\,\rho} + m_2\,c_x^2\,e^{2\,i\,\sigma}\right),\nonumber \\
M_{\nu\;33} &= & s_y^2\,e^{-2\,i\,\d}\,
\left(m_1\,s_x^2\,e^{2\,i\,\rho} + m_2\,c_x^2\,e^{2\,i\,\sigma}\right).
\label{m3zeroS}
\eea
The mixing angles $\theta_x$ and
$\theta_y$ are constrained by data as given in
eq.~(\ref{expdata}). However, there is no constraint on the phases
$\d , \rho$ and $\sigma$, thus for such a model there is no
definite prediction for these phases. For the sake of
illustration, we present a numerical value for the mass matrix for
$\d=75^0 , \rho=30^0$ and $\sigma=30^0$, while $\theta_x$ and
$\theta_y$ take their central value as given in
eq.~(\ref{expdata}), \be M_\nu = m_1\,\left(
\begin {array}{ccc}
0.497+0.862\,i&-0.006+0.001\,i& 0.005-0.001\,i\\
-0.006+0.001\,i& 0.106 \times 10^{-9}-0.546\,i& -0.123\times 10^{-9}+0.492\,i\\
0.005-0.001\,i& -0.123\times 10^{-9}+0.492\,i& 0.909\times 10^{-10}-0.443\,i
 \end {array}
\right).
\label{m3zeroS}
\ee
%%%%%%%%%%%%%%%%%%%%%%%%%%%%%%%%%%%%%%%%%%%%%%%%%%%%%%%%%%%%%%%%%%%%%%%%%%%%%
%%%%%%%%%%%%%%%%%%%%%%%%%%%%%%%%%%%%%%%%%%%%%%%%%%%%%%%%%%%%%%%%%%%%%%%%%%%%%

%%%%%%%%%%%%%%%%%%%%%%%%%%%%%%%%%%%%%%%%%%%%%%%%%%%%%%%%%%%%%%%%%%%%%%
%%%%%%%%%%%%%%%%%%%%%%%%%%%%%%%%%%%%%%%%%%%%%%%%%%%%%%%%%%%%%%%%%%%%%%

\section{Failing cases}
{\it Pattern} {\bf F1, F2, F3, F4} and {\bf F5}:
The patterns  {\bf F1, F2} and {\bf F3} are not acceptable since they lead to $m_2 < m_1$ , where as the patterns {\bf F4} and {\bf F5}
 can not account for the neutrino
oscillation since they give $m_1=m_2=m_3$.

\begin{landscape}
\begin{table}[h]
\begin{center}
{%\tiny
\begin{tabular}{ c|c c c c c c c c c c c c c c c c}
\hline
\hline\\
   \mbox{Model} &\mbox{Minors} & Hierarchy &  $\th_x$   & $\th_y$ & $\th_z$  & $\d$
   & $R_\n$ & $\frac{m_1}{m_3}$ & $\frac{m_2}{m_3}$ & $\rho$& $\sigma$
   &$m_3$&$\Delta m^2_{\mbox{atm}}$&$\Sigma$ &$\langle m \rangle_{ee}$
   & $\langle m \rangle_{e}$ \\  \\
\hline\\
${\bf A1}$ & $33,32$ & N &$34$ &  $42$ & $7$   & $0$
& $0.0298$ & $0.060$ & $0.180$ & $0$ & $90$ & $0.052$ &
$0.0027$ & $0.065$ &  $0$ & $0.009$   \\
${\bf A2}$ & $22,32$ & N &${\tiny //}$ &  $48$ & ${\tiny //}$   & $180$
& ${\tiny //}$ & ${\tiny //}$ & ${\tiny //}$ & ${\tiny //}$ & ${\tiny //}$ & ${\tiny //}$ &
${\tiny //}$ & ${\tiny //}$ &  ${\tiny //}$ & ${\tiny //}$   \\
 ${\bf B3}$ & $33,31$ & N & $34$ &  $42$ & $5$   &
$88.3$ & $0.0325$ & $0.807$ & $0.814$ & $179$ & $178$ & $0.085$ &
$0.0025$ & $0.223$ & $0.069$ & $0.069$\\
 ${\bf B4}$ & $22,21$ & N & ${\tiny //}$ &  $48$ & ${\tiny //}$   &
$268.3$ & ${\tiny //}$ & ${\tiny //}$ & ${\tiny //}$ & ${\tiny //}$ & ${\tiny //}$ & ${\tiny //}$ &
${\tiny //}$ & ${\tiny //}$ & ${\tiny //}$ & ${\tiny //}$\\
${\bf B4}$ & $22,12$ & I & $34$ &  $42$ & $5$ &
$90.4$ & $0.0310$ & $1.227$ & $1.234$ & $1.16$ & $0.05$ &$0.070$&
$0.0026$& $0.242$& $0.086$
& $0.086$ \\
${\bf B3}$ & $33,31$ & I & ${\tiny //}$ &  $48$ & ${\tiny //}$ &
$270.4$ & ${\tiny //}$ & ${\tiny //}$ & ${\tiny //}$ & ${\tiny //}$ & ${\tiny //}$ &${\tiny //}$&
${\tiny //}$& ${\tiny //}$& ${\tiny //}$
& ${\tiny //}$ \\
${\bf B5}$ & $33,12$  & N& $34$  & $42$ & $5$   &
$92$ & $0.0337$ & $0.802$ & $0.809$ & $1.4$ & $2.34$ & $0.082$ &
$0.0024$ & $0.215$ &  $0.066$ & $0.066$
\\
${\bf B6}$ & $22,13$  & N& ${\tiny //}$  & $48$ & ${\tiny //}$   &
$272$ & ${\tiny //}$ & ${\tiny //}$ & ${\tiny //}$ & ${\tiny //}$ & ${\tiny //}$ & ${\tiny //}$ &
${\tiny //}$ & ${\tiny //}$ &  ${\tiny //}$ & ${\tiny //}$
\\
${\bf B6}$ & $22,13$ & I & $34$ &  $42$ & $5$ &
$89.83$ & $0.033$ & $1.234$ & $1.241$ & $179$ & $0.26$ &$0.066$&
$0.0024$& $0.231$& $0.082$
& $0.082$ \\
${\bf B5}$ & $33,12$ & I & ${\tiny //}$ &  $48$ & ${\tiny //}$ &
$269.83$ & ${\tiny //}$ & ${\tiny //}$ & ${\tiny //}$ & ${\tiny //}$ & ${\tiny //}$ &${\tiny //}$&
${\tiny //}$& ${\tiny //}$& ${\tiny //}$
& ${\tiny //}$ \\

${\bf D}$ & $22,33$ & N &$34$ &  $41$ & $5$ & $52.7$
& $0.033$ & $0.512$ & $0.534$ & $14.35$ & $128.95$ &$0.058$& $0.0024$&
$0.118$& $0.016$
& $0.030$ \\

\hline
\hline
\end{tabular}
}
\end{center}
 \caption{\small  The acceptable patterns for the
 two-vanishing minors. The minor corresponding to the index $(ij)$
 is the determinant of the sub-matrix obtained by deleting the
 $i^{th}$ line and the $j^{th}$ column. All the angles (masses) are
 evaluated in degrees ($eV$) (the mark $//$ means a value equal to what overheads it).}
\label{tab1}
 \end{table}

\end{landscape}
%%%%%%%%%%%%%%%%%%%%%%%%%%%%%%%%%%%%%%%%%%%%%%%%%%%%%%%%
%%%%%%%%%%%%%%%%%%%%%%%%%%%%%%%%%%%%%%%%%%%%%%%%%%%%%%%%

\section{Symmetry realization}
All textures with  zero minors discussed in this work can be realized in a
simple way in models based on seesaw mechanism with a flavour Abelian
symmetry. Zero minors in the non singular neutrino mass matrix ($M_\nu$) are
equivalent to zeros in the inverse mass matrix $(M_\nu^{-1})$. This
equivalence, however, is valid only for invertible mass matrices. In turn,
eq.~(\ref{see-saw}), tells us that, in case of diagonal neutrino Dirac
mass matrices, zeros in $(M_\nu^{-1})$ lead to zeros
in $(M_R)$.

In order to construct the required leptonic mass matrices, we need three
right-handed neutrinos $\nu_{Rj}$, three right-handed charged
leptons $l_{Rj}$ and three left-handed lepton doublets
$D_{Lj}=(\nu_{Lj} , l_{Lj})^{T}$, where $j$ is the family index running from $1$
to $3$. As to the scalar sector, one Higgs doublet, the standard
model SM Higgs, and many scalar singlets are required.

The underlying symmetry for building up mass matrices can be an Abelian discrete one
as invoked in~\cite{Lavoura}. Following the same strategy, one can
utilize $Z_8$ for constructing the model ${\bf B3}$ which is
characterized by vanishing minors $(33,31)$, and hence zeros at positions $(33,31)$ in $(M_\nu^{-1})$.
 Under the action of $Z_8$, leptons of the
first family remain invariant, those of the second family change
sign, and those of the third get multiplied by $\omega=\exp{({i\,\pi\over
4})}$, whereas the SM Higgs doublet remains
invariant. These assigned transformations under $Z_8$ automatically
generate diagonal Dirac mass matrices for both charged leptons
and neutrinos.

The bilinears $\nu_{Ri} \nu_{Rj}$ under $Z_8$, necessary for constructing the Majorana
mass for right handed neutrino, transform as
\be
\left(
\begin {array}{ccc}
1&\omega^4& \omega\\
\omega^4& 1& \omega^5\\
\omega&\omega^5&\omega^2
\end {array}
\right),
\label{N1bi}
\ee
The $(1,1)$ and $(2,2)$ matrix elements of $M_R$ are $Z_8$
invariant, hence their corresponding mass terms are directly present in the
Lagrangian. The $(1,2)$ matrix element requires the presence of a
real scalar singlet (call it $\chi_{12}$) which changes sign under $Z_8$. The $(2,3)$
matrix element is generated by the Yukawa coupling to a complex
scalar singlet (call it $\chi_{23}$) which gets multiplied by $\omega^3$ under $Z_8$.
The other entries of $M_R$ remains zero in the absence of any
further scalar singlets. The resulting right-handed Majorna mass
matrix can be casted in the form,
\be
\left(
\begin {array}{ccc}
\times &\times& 0\\
\times& \times& \times\\
0&\times&0
\end {array}
\right),
\label{N1MR}
\ee
where the cross sign denotes a non-vanishing element.

%As to the model ${\bf N3}$, it can be generated by utilizing $Z_4$
%symmetry. Under the action of $Z_4$, leptons of the
%first and third families get multiplied by $i$ and $-i$ respectively, while
%those of the second family as well as the SM Higgs remain invariant.
% As in case ${\bf N1}$, the assigned transformations
%guarantee diagonal Dirac mass matrices for both charged leptons
%and neutrinos, and hence zeros in $M_\nu^{-1}$ would appear as zeros in $M_R$.
%The bilinears $\nu_{Ri} \nu_{Rj}$ under $Z_4$ transform as
%\be
%\left(
%\begin {array}{ccc}
%-1&i& 1\\
%i& 1& -i\\
%1&-i&-1
%\end {array}
%\right).
%\label{N3bi}
%\ee
%The $(2,2)$ and $(1,3)$ matrix elements of the symmetric $M_R$ are $Z_4$
%invariant, hence their corresponding mass terms are directly present in the Lagrangian.
%Only one complex scalar singlet (call it $\chi_{12}$) which gets multiplied by $-i$ is
%needed together with its complex conjugate (call it $\chi_{32}$) to generate
%the entries $(1,2)$ and $(3,2)$ in $M_R$. The resulting mass matrix $M_R$ takes
%the form
%\be
%\left(
%\begin {array}{ccc}
%0 &\times& \times\\
%\times& \times& \times\\
%\times&\times&0
%\end {array}
%\right),
%\label{N3MR}
%\ee
%where the cross sign denotes again a non-vanishing element.

Out of fifteen possible models, nine of them ${\bf A1}$, ${\bf A2}$, ${\bf B3}$, ${\bf B4}$,
${\bf B5}$, ${\bf B6}$, ${\bf S1}$, ${\bf S2}$
and ${\bf F3}$ can be constructed using $Z_8$
symmetry. The leftover models can be generated using $Z_4$
symmetry. Symmetry realizations for all models are summarized in
Table.~(\ref{tab3}), where the transformation properties for each
lepton family and needed scalar singlets are given for each model (for completeness, we stated even the phenomenologically
non-successful models).
%The realization obtained for ${\bf N1}$, ${\bf N2}$, ${\bf N4}$, ${\bf N5}$,
%${\bf N8}$, ${\bf I1}$ and ${\bf I3}$ models agree with those
%obtained in~\cite{Lavoura}.
\begin{table}[h]
\begin{center}
\begin{tabular}{|c||c||c|c|c|c|c|c|c|c|c|}
\hline
\multicolumn{11}{|c|}{ $Z_8$ \mbox{Models}}\\
\hline
   \mbox{Model} &\mbox{Minors} & $1_F $ &$2_F $   & $3_F $ & $\chi_{11}$  &
   $\chi_{12}$& $\chi_{13}$ & $\chi_{22}$ & $\chi_{23}$ & $\chi_{33}$  \\
\hline
 ${\bf B3}$ & $33,31$ & $1$ & $-1$ &  $\omega$ & absent   &
$-1$ & absent & absent & $\omega^3$ & absent
\\
\hline
 ${\bf B5}$ & $33,12$ & $1$ & $-1$ &  $\omega$ & absent   &
 absent & $\omega^7$ & absent & $\omega^3$ & absent\\
 \hline
 ${\bf A1}$ & $33,32$ & $1$ & $-1$ &  $\omega$ & absent   &
 $-1$ & $\omega^7$ & absent & absent & absent\\
\hline
 ${\bf F3}$ & $11,32$ & $\omega$ & $-1$ &  $1$ & absent   &
 $\omega^3$ & $\omega^7$ & absent & absent & absent\\
\hline
 ${\bf A2}$ & $22,32$ & $1$ & $\omega$ &  $-1$ & absent   &
 $\omega^7$ & $-1$ & absent & absent & absent\\
\hline
${\bf B4}$ & $22,12$ & $1$ & $\omega$ &  $-1$ & absent   &
 absent & $-1$ & absent & $\omega^3$ & absent\\
 \hline
 ${\bf S1}$ & $31,11$ & $\omega$ & $-1$ &  $1$ & absent   &
 $\omega^3$ & absent & absent & $\omega^4$ & absent\\
 \hline
 ${\bf B6}$ & $22,31$ & $1$ & $\omega$ &  $-1$ & absent   &
 $\omega^7$ & absent & absent & $\omega^3$ & absent\\
 \hline
 ${\bf S2}$ & $11,12$ & $\omega$ & $-1$ &  $1$ & absent   &
 absent & $\omega^7$ & absent & $\omega^4$ & absent\\
\hline
\multicolumn{11}{|c|}{ $Z_4$ \mbox{Models}}\\
\hline
${\bf F1}$ & $33,11$ & $i$ & $1$ &  $-i$ & absent   &
 $-i$ & absent & absent & $i$ & absent\\
\hline
${\bf D}$ & $22,33$ & $1$ & $i$ &  $-i$ & absent   &
 $-i$ & $i$ & absent & absent & absent\\
\hline
${\bf F2}$ & $22,11$ & $i$ & $-i$ &  $1$ & absent   &
 absent & $-i$ & absent & $i$ & absent\\
\hline
${\bf F4}$ & $31,32$ & $i$ & $-i$ &  $1$ & $-1$   &
 absent & absent & $-1$ & absent & absent\\
\hline
${\bf S3}$    & $31,12$ & $1$ & $i$ &  $-i$ & absent   &
 absent & absent & $-1$ & absent & $-1$\\
\hline
 ${\bf F5}$ & $32,12$ & $i$ & $1$ &  $-i$ & $-1$   &
 absent & absent & absent & absent & $-1$\\
\hline
\end{tabular}
\end{center}
 \caption{\small  The symmetry realization for $15$ patterns of
 two-vanishing minors. The index $1_F$ indicates the lepton first
 family and so on. The $\chi_{kj}$ denotes a scalar singlet which produce
 the entry $(k,j)$ of the right-handed Majorana mass matrix when acquiring a vev
  at the see-saw scale. The transformation properties, under the
 specified group, is listed below each lepton family and needed scalar singlets for each
 model. $\omega$ denotes $\exp{({i\,\pi\over
4})}$, while $i=\sqrt{-1}$. }
\label{tab3}
 \end{table}

The question arises whether or not this seesaw enforcement of zero minors at high scale persists down to low scales.
Actually, the renormalization group analysis shows that this is possible with one  Higgs doublet \cite{Lavoura}, since
the matrices $M_\n$ at any two energy scales $\mu_1$ and $\mu_2$
are related by~\cite{chankowski}
\be
M_\n \left( \mu_1 \right) = I M_\n \left( \mu_2 \right) I,
\label{relation}
\ee
where the matrix $I$(which depends on $\mu_1$ and $\mu_2$)
is diagonal, positive, and non-singular.
It follows that any zero minor in  $M_\n$, at a given energy scale,
remains zero at any other energy scale.

The models ${\bf S1}$, ${\bf S2}$ and $\bf S3$ in the limit of vanishing
$m_3$ and $\theta_z$ together give a singular mass matrix $M_\nu$. To
produce such models of acceptable non-invertible textures within seesaw schemes, $M_R$ can not be singular otherwise the sea-saw
mechanism would not work, and hence the only choice is to have a singular
Dirac neutrino mass matrix. A simple guess, which by no means
excludes other possibilities, is where $M_R$ and $M_D$ have the following
forms:
\be
\begin{array}{cc}
M_R=\left(
\begin {array}{ccc}
0&\alpha& 0\\
\alpha &\beta & 0\\
0& 0 & \gamma
\end {array}
\right),
&
M_D=\left(
\begin {array}{ccc}
D&0& A\\
0 &0 & B\\
0& 0 & C
\end {array}
\right),
\end{array}
\label{singRD}
\ee
where $A, B, C, D, \alpha, \beta$ and $\gamma$ are arbitrary
parameters.
The resulting neutrino mass matrix generated through seesaw
mechanism takes the form,
\be
M_\nu=\left(
\begin {array}{ccc}
-{D^2 \beta \over \alpha^2} + {A^2\over \gamma} &{A B\over \gamma}& {A C\over \gamma}\\
{A B\over \gamma}& {B^2\over \gamma}& {B C\over \gamma}\\
{A C\over \gamma}&{B C\over \gamma}&{C^2\over \gamma}
\end {array}
\right).
\label{singMn}
\ee
This matrix has a zero eigenvalue $(m_3=0)$ and vanishing  minors $(31,11,12)$ as required.
The symmetry realization for this construction could be done
through a generic choice of the group $Z_{12}\times Z_2$ discussed
in~\cite{Grimus}, which may not be the most economic way. The
leptonic fields transform as
\be
\begin{array}{ccc}
l_{R1}\rightarrow \theta l_{R1}, & \nu_{R1}\rightarrow \theta \nu_{R1}, &
\overline{D}_{L1} \rightarrow \theta \overline{D}_{L1},\\
l_{R2}\rightarrow \theta^2 l_{R2}, & \nu_{R2}\rightarrow \theta^2 \nu_{R1}, &
\overline{D}_{L2} \rightarrow \theta^3 \overline{D}_{L2},\\
l_{R3}\rightarrow \theta^5 l_{R3}, & \nu_{R3}\rightarrow \theta^5 \nu_{R3}, &
\overline{D}_{L3} \rightarrow \theta^8 \overline{D}_{L3},\\
\end{array}
\ee
where $\theta=e^{\frac{i\pi}{6}}$. Hence, the bilinears $\overline{D}_{Lj}\,l_{Rk}$, $\overline{D}_{Lj}\,\nu_{Rk}$
and $\nu_{Rj}\,\nu_{Rk}$ transform as
\be
\begin{array}{cc}
\overline{D}_{Lj}\,l_{Rk}\sim \overline{D}_{Lj}\,\nu_{Rk}\sim\left(
\begin {array}{ccc}
\theta^2&\theta^3& \theta^6\\
\theta^4 &\theta^5 & \theta^8\\
\theta^9& \theta^{10} & \theta
\end {array}
\right),
&
\nu_{Rj}\,\nu_{Rk}\sim \left(
\begin {array}{ccc}
\theta^2&\theta^3& \theta^6\\
\theta^3&\theta^4& \theta^7\\
\theta^6&\theta^7& \theta^{10}\\
\end {array}
\right).
\end{array}
\label{sing_Trans}
\ee
To achieve a diagonal charged lepton mass matrix, only three scalar
Higgs doublets are needed which are denoted by $\Phi_{11}$, $\Phi_{22}$
and $\Phi_{33}$. Under the action of $Z_{12}$ these scalar
doublets get respectively multiplied by $\theta^{10}$, $\theta^{7}$
and $\theta^{11}$. The Dirac and Majorana neutrino mass matrices
in eq.~(\ref{singRD}) can be achieved equally by introducing scalar Higgs fields with suitable transformation properties,
namely four scalar
doublets $\tilde{\Phi}_{11}$, $\tilde{\Phi}_{13}$,
$\tilde{\Phi}_{23}$, and $\tilde{\Phi}_{33}$, being multiplied respectively by $\theta^{10}$, $\theta^{6}$, $\theta^{4}$
and $\theta^{11}$ under $Z_{12}$, for the Dirac mass term,
and three scalar singlets $\chi_{12}$, $\chi_{22}$ and  $\chi_{33}$, multiplied respectively by
$\theta^{9}$, $\theta^{8}$ and $\theta^{2}$ under $Z_{12}$, for the Majorana mass term.
The $\tilde{\Phi}_{jk}$ scalar Higgs
doublet and $\nu_{Rj}$ change sign under $Z_2$, while all other
multiplets remain invariant. It is important to notice that
the scalar Higgs doublets take vacuum expectation values (vev) at the
electro-weak scale, while scalar singlets acquire vevs at the seesaw scale.

\section{ Summary and conclusions}
We studied all the possible patterns of Majorana neutrino mass
matrices with two independent vanishing minors.

For the possible fifteen cases, we found seven patterns ({\bf A1,
A2, B3, B4, B5, B6} and {\bf D}) able to accommodate current data,
without need to tune the input parameters.

Five patterns ({\bf S1, S2,
F1, F2} and {\bf F3}) lead to $m_2$ smaller than $m_1$ and thus are non-acceptable. Three
cases ({\bf S3, F4,
} and {\bf F5}) predict degeneracy of masses $m_1 = m_2 = m_3$, and thus can
not make room for the oscillation phenomena.

The acceptable patterns can be classified into three categories ({\bf A, B} and {\bf D}). The classes {\bf A} and {\bf D} allow only
for N-type hierarchy, whereas the class {\B} has room for both N- and I-type of hierarchies.

%The patterns of class {\bf A} present, numerically,
%approximate real mass matrices, and thus it would be difficult to
%display CP violation phenomena with these patterns.
The six cases
({\bf A1, A2, B3} and {\bf B4}) appear as `two-zero' textures, and the
corresponding analytical expressions agree with
\cite{Xing}. The other three two-zero textures allowed by current
data, and denoted by ({\bf B1, B2} and {\bf C}) in \cite{Xing}, are not
reproduced as two-vanishing minors textures. On the other hand,
there are three two-vanishing minor textures ({\bf B5, B6} and {\bf D}) which
are allowed by data and do not show up as two-zero textures. This
classification of patterns coincides exactly with the results of
\cite{Lavoura}. However, as mentioned in the introduction, the
analytical parts of the two approaches are not completely
identical, so that one needs not to assume an invertible neutrino
mass matrix in the `vanishing minor' approach.

%We took, except for the allowed cases {\bf A4}, {\bf A2} textures,
%the value $5^\circ$ for the angle $\theta_z$. For these two cases
%({\bf A1}, {\bf A2}),  the model starts to be consistent with the
%experimental data when $\theta_z$ ($\ge 7^0$ ,$\ge 6^0$ )
%respectively. The stability of the results against the variation
%of $\theta_z$, wherever it is possible, from nearly $0^0$ to
%$10^0$ is achieved in the models of class {\bf B}.

In all the successful models, the non oscillation parameters $\langle m \rangle_{e}
, \langle m\rangle_{ee}\,\mbox{and}\, \Sigma $ are consistent with the bounds given
in eq.~(\ref{expdata}).
In addition, $\langle m \rangle_{e}, \langle m
 \rangle_{ee}$ and $m_3$ have the same order of magnitude.
The mass sum parameter is always constrained to be  $\Sigma \le
0.242\;\mbox{eV}$ which is safe with the cosmological bound in eq.~(\ref{expdata}).

There are some characteristic features which might differentiate the classes from each other.
First, for the class {\bf A} patterns, there is an acute hierarchy in the mass spectrum since the combination $s_z t_y$ appears
in the mass ratio masses making the hierarchy sharp for small enough $s_z$. The hierarchy in class {\bf B} is less sharp and
the three masses are all of the same order of magnitude, since $t_y$ alone appears in the leading term of the expansion of the mass ratios.
Second, in the class {\bf A} there is a lower bound on $\t_z$, whereas no such bound in classes {\bf B} and {\bf D}.
Third, since the Dirac angle $\d$ is not constrained in the class $A$ or $D$, whereas it is likely to be around right angles in the class {\bf B}
, then it is in this latter class that CP is likely to be maximally violated.
Fourth, there is no signal for neutrinoless double beta decay in the class {\bf A}, while the corresponding parameter $\langle m \rangle_{ee}$
does not vanish in the classes {\bf B} and {\bf D}.

Table (\ref{classB}) presents the key signatures involving ($\d, \t_y$) allowing to distinguish among the patterns of the class {\bf B}.
Also, there is a `correlation' between these two parameters in class {\bf D} in that $\t_y$ being in first (second) octant leads to
 $c_\d > 0$ ($c_\d < 0$), whereas no such relation in class {\bf A}.

New results are obtained in case of non-invertible mass matrix
which can not be recovered in \cite{Lavoura}. Those correspond to
the patterns {\bf S1}, {\bf S2} and {\bf S3}. For these models, we only find a consistent
solution when both $m_3$ and $\theta_z$ vanish, a solution that
accommodates $\theta_x$ and $\theta_y$ to be in their acceptable
range given in eq.~(\ref{expdata}), while  the phases $\d$,
$\sigma$ and $\rho$ are not constrained. Peculiarly enough, the
three cases have identical mass matrix as given in
eq.~(\ref{m3zeroS}), and thus it would be difficult to
distinguish between the three models in this limit of vanishing
$m_3$ and $\theta_z$. An interesting thing about these three
models is that they correspond to two-zero textures before taking
the limit of vanishing $m_3$ and $\theta_z$, whereas they cease to
be so after taking the limit where they become only characterized
by vanishing two minors and the whole determinant.

%five of the allowed models (the remaining
%ones except ${\bf N5}$). In fact, the stability is also achieved
%in two of the disallowed models, namely ${\bf I2}$ and ${\bf I4}$.
%As to the Dirac-phase, we found the solution in many cases,
%especially for the allowed models, to depend sensitively on its
%value, so that any deviation from it violates the experimental
%constraints.
%
%This point was not addressed in \cite{Lavoura}.
%
% and
%we illustrate it in figure (\ref{delta}), corresponding to case
%{\bf I1} of the two-vanishing minors texture. Here, for the choice
%$\theta_x = 34^\circ$, $\theta_y = 42^\circ$ and $\theta_z =
%5^\circ$, we vary $\delta$ and find that satisfying the condition
%(\ref{Rconstraint}) singles out a nearly right angle for $\delta$.
%This offers a good way for selecting the appropriate Dirac phase
%$\d$ which we adopted in our present work.

Concluding remarks are in order. First, for all patterns the
leading terms in the expansion of the angles ($\rho,\sigma$) in
powers of ($\sin(\th_z)$) are either $\d$ or $\frac{\d}{2}$ (mod
$\frac{\pi}{2}$). Second, the models of the classes {\bf B} and {\bf D},
%{\bf N5},
%{\bf I1} and {\bf I3}
do not have room for the atmospheric mixing
angle to be maximal (${\pi\over 4}$), otherwise the parameter
$R_\nu$ would become too large. Third, all models can be generated
in the frame work of flavour Abelian discrete symmetry, and additional scalar fields
with appropriate transformation properties, implemented in seesaw schemes.

\section*{{\large \bf Acknowledgements}}
Both authors would like to thank L. Lavoura for drawing our attention to his work.
One of the authors, E. I. Lashin would like to thank both of A. Smirnov and S. Petcov
for useful discussions. Part of this work was done within the associate scheme of ICTP.

\end{document}